\documentclass[preprintnumbers, onecolumn, floatfix, preprintnumbers, amsmath, amssymb, superscriptaddress]{revtex4}

\usepackage{bm,graphicx,dcolumn,epstopdf,epsf, latexsym,mathbbol, amssymb,amsmath,color,slashed, mathrsfs,mathcomp,simplewick}
\usepackage[center]{subfigure}
\usepackage{multirow}
\usepackage{makecell}
\usepackage[colorlinks,linkcolor=blue,citecolor=blue,urlcolor=blue]{hyperref}

\begin{document}
\allowdisplaybreaks
 \newcommand{\bq}{\begin{equation}}
 \newcommand{\eq}{\end{equation}}
 \newcommand{\bqn}{\begin{eqnarray}}
 \newcommand{\eqn}{\end{eqnarray}}
 \newcommand{\nb}{\nonumber}
 \newcommand{\lb}{\label}
 \newcommand{\f}{\frac}
 \newcommand{\p}{\partial}
\newcommand{\PRL}{Phys. Rev. Lett.}
\newcommand{\PLB}{Phys. Lett. B}
\newcommand{\PRD}{Phys. Rev. D}
\newcommand{\CQG}{Class. Quantum Grav.}
\newcommand{\JCAP}{J. Cosmol. Astropart. Phys.}
\newcommand{\JHEP}{J. High. Energy. Phys.}
\newcommand{\red}{\textcolor{red}}

\title{Testing parity symmetry of gravity with gravitational waves}

\author{Jin Qiao}
\email{qiaojin@zjut.edu.cn}
\affiliation{Purple Mountain Observatory, Chinese Academy of Sciences, Nanjing, 210023, P.R.China}
\affiliation{ School of Astronomy and Space Sciences, University of Science and Technology of China, Hefei, 230026, P.R.China}
\affiliation{ Department of Astronomy, University of Science and Technology of China, Hefei 230026, P.R.China}

\author{Zhao Li}
\affiliation{ School of Astronomy and Space Sciences, University of Science and Technology of China, Hefei, 230026, P.R.China}
\affiliation{ Department of Astronomy, University of Science and Technology of China, Hefei 230026, P.R.China}

\author{Tao Zhu}
\affiliation{ Institute for theoretical physics and Cosmology, Zhejiang University of Technology, Hangzhou, 310032, P.R.China}
\affiliation{ United center for gravitational wave physics (UCGWP), Zhejiang University of Technology, Hangzhou, 310032, P.R.China }

\author{Ran Ji}
\affiliation{ Research School of Physics, Australian National University, Acton, ACT, 2601, Australia}

\author{Guoliang Li}
\affiliation{Purple Mountain Observatory, Chinese Academy of Sciences, Nanjing, 210023, P.R.China}
\affiliation{ School of Astronomy and Space Sciences, University of Science and Technology of China, Hefei, 230026, P.R.China}

\author{Wen Zhao}
\email{wzhao7@ustc.edu.cn}
\affiliation{ School of Astronomy and Space Sciences, University of Science and Technology of China, Hefei, 230026, P.R.China}
\affiliation{ Department of Astronomy, University of Science and Technology of China, Hefei 230026, P.R.China}







\date{\today}

\begin{abstract}
The examination of parity symmetry in gravitational interactions has drawn increasing attention. Although Einstein's General Relativity is parity-conserved,
numerous theories of parity-violating (PV) gravity in different frameworks have recently been proposed for different motivations. In this review, we briefly summarize the recent progress of these theories, and focus on the observable effects of PV terms in the gravitational waves (GWs), which are mainly reflected in the difference between the left-hand and right-hand polarization modes. We are primarily concerned with the implications of these theories for GWs generated by the compact binary coalescences and the primordial GWs generated in the early Universe. The deviation of GW waveforms and/or primordial power spectrum can always be quantified by the energy scale of parity violation of the theory. Applying the current and future GW observation from laser interferometers and cosmic microwave background radiation, the current and potential constraints on the PV energy scales are presented, which indicates that the parity symmetry of gravity can be tested in high energy scale in this new era of gravitational waves.


\end{abstract}

\maketitle

\section{Introduction}
\renewcommand{\theequation}{1.\arabic{equation}} \setcounter{equation}{0}

The application of gravitational waves (GWs) to test gravitational theories becomes an important topic in gravitational wave astronomy. This review mainly focuses on the theories of parity-violating gravity and their applications to different scenarios in astronomy. Theoretically, a variety of parity-violating gravities have been constructed, all of which have been extensively studied. These parity-violating gravities formally fall into two main categories: One is based on the framework of Riemannian geometry with modifications to the Einstein-Hilbert action to obtain parity violation, and the other is in the framework of non-Riemannian geometry where the alternative gravity of general relativity is modified to produce parity violation. In the framework of Riemannian geometry, the classical type of gravity is Chern-Simons (CS) gravity \cite{Jackiw:2003pm}, which modifies GR by introducing the CS term. {CS gravity is further extended to the most general parity-violating scalar-tensor gravity by introducing coupling terms for the higher order derivatives of the scalar field \cite{Crisostomi:2017ugk}. The problem of parity violation in the Poincar\'{e} gauge theory of gravity has been analyzed, where models are built as natural extensions of Einstein-Cartan theory \cite{Obukhov:2020zal}.} Furthermore, by breaking the time diffeomorphism (or Lorentz symmetry), one can naturally introduce parity-violating but spatially covariant terms into the action, thus forming Ho\v{r}ava-Lifshitz (HL) gravity \cite{Wang:2012fi, Takahashi:2009wc, Zhu:2013fja} and spatially covariant gravity \cite{Gao:2019liu, Gao:2020yzr}.

The most classical theories of gravity that have been developed in a non-Riemannian geometric framework are teleparallel gravities (TGs) \cite{BookTG} and symmetric teleparallel gravity (STG) \cite{Nester:1998mp}. TG is gravity described in terms of spacetime torsion rather than curvature, where the model equivalent to GR is called the Teleparallel Equivalent of General Relativity (TEGR). TEGR gravity is modified by the introduction of an odd parity-violating topological term consisting of torsion, i.e.  Nieh-Yan (NY) modified gravity \cite{Li:2020xjt, Li:2021wij}. STG theory is formulated in spacetime given zero curvature and zero torsion and attributes gravity to a non-metricity, where the model equivalent to GR is called the Symmetric Teleparallel General Relativity (STGR). The STGR can also be modified by a parity-violating term to form parity-violating symmetric teleparallel gravity \cite{Conroy:2019ibo}.

Observationally, the presence of parity violation in GWs produced by isolated sources affects the waveform of GWs propagation in two ways. One way is to modify the conventional dispersion relation for GWs. This causes the velocities of the left- and right-hand circular polarization of the GWs to be different, i.e. the velocity birefringence of GWs. This phenomenon has been found to exist in HL gravity \cite{Zhao:2019xmm}, {Chern-Simons Axion Einstein Gravity \cite{Nojiri:2019nar}, Chern-Simons Axion $F$(R) Gravity \cite{Nojiri:2020pqr},} NY modified teleparallel gravity \cite{Li:2020xjt, Wu:2021ndf}, and parity-violating symmetric teleparallel gravity \cite{Conroy:2019ibo, Li:2021mdp}. Another way of parity violation is to change the frictional terms in the GWs propagation equation, where these additional frictional terms modify the amplitude of the GWs. Thus the amplitude of left-hand circular polarization of GWs increases (or decreases) during the propagation, while the amplitude for the right-hand mode decreases (or increases), i.e. the amplitude birefringence of GWs.  The correction of this phenomenon to the GW waveform has been studied in the framework of CS modified gravity in \cite{Yunes:2010yf, Yagi:2012ya, Alexander:2017jmt, Yagi:2017zhb,Li:2022grj}.
{Recently, the research in metric-affine CS gravity has found that the metric tensor modes are coupled to the torsion tensor components, leading to the appearance of velocity birefringence \cite{Boudet:2022nub, Bombacigno:2022naf}. The studies in Palatini CS  gravity have shown that both amplitude and velocity birefringence effects are present in the propagation of GWs polarization \cite{Sulantay:2022sag}.}
It also has been shown in the parity-violating scalar-tensor gravity that parity violation leads to the presence of both phenomena in the waveforms of GWs \cite{Nishizawa:2018srh, Gao:2019liu, Qiao:2019wsh, Zhao:2019xmm}.

Experimentally, as the sensitivity of the Advanced LIGO, Virgo and KAGRA detector continues to be upgraded, detected GWs events accumulate at an increasing rate \cite{LIGOScientific:2018mvr,LIGOScientific:2020ibl,LIGOScientific:2021usb,LIGOScientific:2021djp}. With the catalog of detected events, like GR \cite{Will:1997bb}, a variety of more accurate tests of parity-violating gravities can be performed. The most extensive research has been carried out to search for birefringence effects in the propagation of GWs in GW data.
Studies based on this are divided into theory-independent approaches \cite{Shao:2020shv}, of which the method on mode splitting in \cite{Zhao:2019szi}, Bayesian analysis performed in \cite{Wang:2020cub, Wang:2021gqm, huqian, Zhao:2022pun, Wang:2020pgu}, and Fisher matrix analysis considered in \cite{Wang:2017igw}, and model-dependent approaches, of which Chern-Simons gravity is constrained in \cite{Okounkova:2021xjv}, Nieh-Yan modified teleparallel gravity considered in \cite{Wu:2021ndf}, spatial covariant gravity analyzed in \cite{Zhu:2022uoq}, and more generic parity and Lorentz violating gravities investigated in \cite{Gong:2021jgg,Niu:2022yhr}.

Different from gravitational waves produced by the isolated sources, primordial gravitational waves (PGWs) come from quantum fluctuations and carry important information about the early universe, such as the physics of inflation, bouncing and emergent universe. The most effective way to detect PGWs is to measure the B-mode polarization of the cosmic microwave background (CMB).  In CMB, the PGWs can produce the TT, EE, BB, and TE spectra, but the TB and EB spectra vanish if the parity symmetry in gravity is respected \cite{Krauss:2010ss, Garcia-Bellido:2010uka, Seljak:1996gy, Kamionkowski:1996zd,zhao2009a,zhao2009b,zhao2014a,zhao2014b}. Since nonzero TB and EB spectra of CMB in large scale implies parity violation in the gravitational sector, the precise measurement of the low-multipole TB and EB spectra could be important evidence of the parity violation of the gravity \cite{Seto:2007tn, Lue:1998mq, Saito:2007kt, QUIET:2010cyv, Gluscevic:2010vv,Ali2022, ali2}.
Towards this purpose, Ref. \cite{Lue:1998mq} first proposed a cosmological study of CS corrected gravity as a way to search for parity-violating effects from the GW sector of the CMB polarization spectrum. Subsequently, with the continuous development of various parity-violating gravities, similar studies were explored in the broader frameworks, such as dynamical Chern-Simons gravity \cite{Peng:2022ttg}, {Chern-Simons $f$(R) Gravity \cite{Odintsov:2022hxu},} parity-violating scalar-tensor gravity \cite{Qiao:2019hkz}, HL gravity \cite{Wang:2012fi, Zhu:2013fja},  spatial covariant gravity \cite{Zhu:2022dfq}, Nieh-Yan modified teleparallel gravity \cite{Li:2020xjt}, and parity-violating symmetric teleparallel gravity \cite{Li:2021mdp}, etc.

In this article, we will briefly review the recent progress on the tests of gravity with gravitational waves.
The paper is organized as follows. In the next section, we briefly introduce the different parity-violating gravities. In Sec. \ref{sec3}, we present the applications of parity-violating gravities to GWs generated by isolated sources. In Sec. \ref{sec4}, we discuss the applications of these parity-violating gravities in the early universe. In Sec. \ref{sec5}, we summarise the conclusions arising from these parity-violating gravities in two applications.

Throughout this paper, the metric convention is chosen as $(-, +, +, +)$, and greek indices $(\mu, \nu, \cdot\cdot\cdot)$ run over $0, 1, 2, 3$ and the latin indices $(i, j, k, \cdot\cdot\cdot)$ run over $1, 2, 3$. We set the units to $c = \hbar=1$.

\section{The parity-violating gravities \label{sec2}}
\renewcommand{\theequation}{2.\arabic{equation}} \setcounter{equation}{0}

In this section, we give a brief review of several parity-violating gravities. We first present the most classical and the simplest parity-violating gravity in the Riemannian geometric framework, namely the CS gravity, followed by its generalization to general parity-violating scalar-tensor gravity. Then the parity-violating Ho\v{r}ava-Lifshitz gravity in the Riemann framework is described. Finally, Nieh-Yan modified teleparallel gravity and parity-violating symmetric teleparallel gravity in non-Riemannian geometry are introduced.

\subsection{ Chern-Simons gravity}
\renewcommand{\theequation}{2.1.\arabic{equation}} \setcounter{equation}{0}

The action of the CS gravity can be written in the following form
\bqn\lb{action}
S &=& \frac{1}{16\pi G} \int d^4 x \sqrt{-g}(R+\mathcal{L}_{\rm CS} + \mathcal{L}_{\phi} + \mathcal{L}_{\rm other}),
\eqn
where $R$ is the Ricci scalar, $\mathcal{L}_{\rm CS}$ is the CS Lagrangian, $\mathcal{L}_\phi$ is the Lagrangian for a scalar field, which may be coupled non-minimally to gravity, and $\mathcal{L}_{\rm other}$ denotes other matter fields. As one of the simplest examples, we consider the action of the scalar field
\bqn
\mathcal{L}_\phi =   \frac{1}{2} \beta g^{\mu \nu} \partial_\mu \phi \partial_\nu \phi + \beta V(\phi).
\eqn
Here $V(\phi)$ denotes the potential of the scalar field. The Lagrangian of CS reads \cite{Alexander:2009tp}
\bqn
\mathcal{L}_{\rm CS} = \frac{\alpha}{4}\vartheta(\phi) \;^*R R,
\eqn
where
\bqn
\;^*R R=\frac{1}{2} \varepsilon^{\mu\nu\rho\sigma} R_{\rho\sigma \alpha\beta} R^{\alpha \beta}_{\;\;\;\; \mu\nu}
\eqn
is the Pontryagin density with $\varepsilon^{\rho \sigma \alpha \beta}$ the Levi-Civit\'{a} tensor defined in terms of the antisymmetric symbol $\epsilon^{\rho \sigma \alpha \beta}$ as $\varepsilon^{\rho \sigma \alpha \beta}=\epsilon^{\rho \sigma \alpha \beta}/\sqrt{-g}$ and the CS coupling coefficient $\vartheta(\phi)$ being an arbitrary function of $\phi$. The parameters $\alpha$ and $\beta$ are coupling constants whose values represent the dynamical ($\alpha \neq 0 \neq \beta$) and non-dynamical ($\alpha \neq0,~\beta =0$) versions of CS gravity. These two versions have no effect on the study of GWs, and in this paper we take $\alpha =1=\beta$.
CS modified gravity effectively extends GR that captures gravitational parity-violating terms in leading order. The similar versions of this theory were suggested in the context of string theory \cite{Campbell:1990fu,Campbell:1992hc}, and three-dimensional topological massive gravity \cite{Deser:1982vy, Deser:1981wh}. However, this theory has a higher-derivative field equation, which induces the dangerous Ostrogradsky ghosts. For this reason, CS modified gravity can only be treated as a low-energy truncation of a fundamental theory. To cure this problem, the extension of CS gravity by considering the terms which involve the derivatives of a scalar field is recently proposed in \cite{Crisostomi:2017ugk} and shown in the next subsection.

\subsection{ The ghost-free parity-violating gravity}
\renewcommand{\theequation}{2.2.\arabic{equation}} \setcounter{equation}{0}

The ghost-free parity-violating gravity is an extension of the CS gravity to the more general parity-violating gravity, where new parity-violating terms $\mathcal{L}_{\rm PV1}$ and $\mathcal{L}_{\rm PV2}$ are introduced into the action (\ref{action}) of the CS gravity.
 $\mathcal{L}_{\rm PV1}$ is the Lagrangian containing the first derivative of the scalar field, which is given by\cite{Crisostomi:2017ugk}
\bqn\lb{LMPV1}
\mathcal{L}_{\rm PV1} &=& \sum_{\rm A=1}^4  a_{\rm A}(\phi, \phi^\mu \phi_\mu) L_{\rm A},\label{lv1}\\
L_1 &=& \varepsilon^{\mu\nu\alpha \beta} R_{\alpha \beta \rho \sigma} R_{\mu \nu}{}^{\rho}{}_{\lambda} \phi^\sigma \phi^\lambda,\nonumber\\
L_2 &=&  \varepsilon^{\mu\nu\alpha \beta} R_{\alpha \beta \rho \sigma} R_{\mu \lambda }^{\; \; \;\rho \sigma} \phi_\nu \phi^\lambda,\nonumber\\
L_3 &=& \varepsilon^{\mu\nu\alpha \beta} R_{\alpha \beta \rho \sigma} R^{\sigma}_{\;\; \nu} \phi^\rho \phi_\mu,\nonumber\\
L_4 &=&  \varepsilon^{\mu\nu\rho\sigma} R_{\rho\sigma \alpha\beta} R^{\alpha \beta}_{\;\;\;\; \mu\nu} \phi^\lambda \phi_\lambda,\nonumber
\eqn
with $\phi^\mu \equiv \nabla^\mu \phi$, and $a_{\rm A}$ are a priori arbitrary functions of $\phi$ and $\phi^\mu \phi_\mu$. In order to avoid the Ostrogradsky modes in the unitary gauge (where the scalar field depends on time only), it is required that $4a_1+2 a_2+a_3 +8 a_4=0$. With this condition, the Lagrangian in Eq.(\ref{lv1}) does not have any higher order time derivative of the metric, but only higher order space derivatives.

One can also consider the terms which contain second derivatives of the scalar field. Focusing on only these that are linear in Riemann tensor and linear/quadratically in the second derivative of $\phi$, the most general Lagrangian $\mathcal{L}_{\rm PV2}$ is given by \cite{Crisostomi:2017ugk}
\bqn\lb{LMPV2}
\mathcal{L}_{\rm PV2} &=& \sum_{\rm A=1}^7 b_{\rm A} (\phi,\phi^\lambda \phi_\lambda) M_{\rm A},\\
M_1 &=& \varepsilon^{\mu\nu \alpha \beta} R_{\alpha \beta \rho\sigma} \phi^\rho \phi_\mu \phi^\sigma_\nu,\nonumber\\
M_2 &=& \varepsilon^{\mu\nu \alpha \beta} R_{\alpha \beta \rho\sigma} \phi^\rho_\mu \phi^\sigma_\nu, \nonumber\\
M_3 &=& \varepsilon^{\mu\nu \alpha \beta} R_{\alpha \beta \rho\sigma} \phi^\sigma \phi^\rho_\mu \phi^\lambda_\nu \phi_\lambda, \nonumber\\
M_4 &=& \varepsilon^{\mu\nu \alpha \beta} R_{\alpha \beta \rho\sigma} \phi_\nu \phi_\mu^\rho \phi^\sigma_\lambda \phi^\lambda, \nonumber\\
M_5 &=& \varepsilon^{\mu\nu \alpha \beta} R_{\alpha \rho\sigma \lambda } \phi^\rho \phi_\beta \phi^\sigma_\mu \phi^\lambda_\nu, \nonumber\\
M_6 &=& \varepsilon^{\mu\nu \alpha \beta} R_{\beta \gamma} \phi_\alpha \phi^\gamma_\mu \phi^\lambda_\nu \phi_\lambda, \nonumber\\
M_7 &=& (\nabla^2 \phi) M_1,\nonumber
\eqn
with $\phi^{\sigma}_\nu \equiv \nabla^\sigma \nabla_\nu \phi$. Similarly, in order to avoid the Ostrogradsky modes in the unitary gauge, the following conditions should be imposed: $b_7=0$, $b_6=2(b_4+b_5)$ and $b_2=-A_*^2(b_3-b_4)/2$, where $A_*\equiv \dot{\phi}(t)/N$ and $N$ is the lapse function. Here, we consider a general scalar-tensor theory with parity violation, which contains all the terms mentioned above. So, the parity-violating terms are given by
\bqn
\mathcal{L}_{\rm PV} = \mathcal{L}_{\rm CS} + \mathcal{L}_{\rm PV1} + \mathcal{L}_{\rm PV2}.
\eqn
The coefficients $\vartheta$, $a_{\rm A}$ and $b_{\rm A}$ depend on the scalar field $\phi$ and its evolution. Therefore, the final action of ghost-free parity-violating gravity is given by
\bqn\lb{xaction}
S &=& \frac{1}{16\pi G} \int d^4 x \sqrt{-g}(R+\mathcal{L}_{\rm PV} + \mathcal{L}_{\phi} + \mathcal{L}_{\rm other}).
\eqn

\subsection{ The parity-violating Ho\v{r}ava-Lifshitz gravity}
\renewcommand{\theequation}{2.3.\arabic{equation}} \setcounter{equation}{0}

The HL gravity is based on the perspective that the Lorentz symmetry appears only as an emergent symmetry at low energies, but can be fundamentally absent at high energies \cite{Horava:2009uw, Wang:2017brl}. This opens a completely new window to build a theory of quantum gravity without the Lorentz symmetry in the UV, using the high-dimensional spatial derivative operators, while still keeping the time derivative operators to the second-order, whereby the unitarity of the theory is reserved. Besides the original version of the theory by Ho\v{r}ava \cite{Horava:2009uw}, there are several modifications, which are absent in several in-consistent problems that appear in the original version. In this paper, we are going to focus on an extension of the HL gravity by abandoning the projectability condition but imposing an extra local U(1) symmetry that was proposed \cite{Zhu:2011xe, Zhu:2011yu}, in which the gravitational sector has the same degree of freedom as that in GR, i.e., only spin-2 massless gravitons exist.

By abandoning the Lorentz symmetry, the HL theory also provides a natural way to incorporate parity violation terms into the theory. For our current purpose, we consider the third- and/or fifth-order spatial derivative operators to the potential term $\mathcal{L}$ of the total action in \cite{Zhu:2013fja, Zhu:2011yu,Zhu:2011xe},
\bqn
\mathcal{L}_{\rm PV} = \f{1}{M^3_{\rm PV}}( \alpha_0 K_{ij} R_{ij} + \alpha_2 \varepsilon ^{ijk} R_{il} \nabla_j R^l_k) + \f{\alpha_1\omega_3(\Gamma)}{M_{\rm PV}} +`` \cdot \cdot \cdot".
\eqn
Here $M_{\rm PV}$ is the energy scale above which the high-order derivative operators become important. The coupling constants $\alpha_0,~\alpha_1, ~\alpha_2$ are dimensionless and arbitrary, $K_{ij}$ and $R_{ij}$ denote, respectively, the extrinsic curvature and the 3-dimensional Ricci tensor built of the 3-metric $g_{ij}$. $\nabla_i$ denotes the covariant derivative with respect to $g_{ij}$, and $\omega_3(\Gamma)$ is the 3-dimensional gravitational CS term. $`` \cdot \cdot \cdot"$ denotes the rest of the fifth-order operators given in Eq.(2.6) of \cite{Zhu:2011yu}. Since they have no contributions to tensor perturbations, in this paper we shall not write them out explicitly.

\subsection{ The Nieh-Yan modified teleparallel gravity}

The teleparallel gravity (TG) theory is a constrained metric-affine theory that is constructed in spacetime with zero curvature and metric compatible connection. The Nieh-Yan modified teleparallel gravity is constructed by introducing parity-violating terms in the GR equivalent teleparallel gravity, which is formulated in flat spacetime with vanishing curvature and vanishing asymmetry.
The action of the Nieh-Yan modified teleparallel gravity is \cite{Li:2020xjt,Wu:2021ndf}
\bqn\lb{NYa}
S 
   &=& \int d^4x \sqrt{-g} \left[ -\f{R(e)}{2}  + \f{c}{4} \theta \mathcal{T}_{A\mu\nu} \mathcal{\widetilde{T}}^{A\mu\nu} + \f{b}{2} \nabla_{\mu}\theta \nabla^{\mu}\theta - bV(\theta) \right] + S_{\rm m},
\eqn
where $b$ is a coupling constant, $\theta$ is a scalar field, and the curvature scalar $R(e)$ is defined by the Levi-Civit\'{a} connection and considered as being constructed entirely from the metric $g_{\mu\nu}$, which in turn is constructed from tetrad field $e^a_{\mu}$, with the relation $g_{\mu\nu} = \eta_{ab}e^a_{\mu}e^b_{\nu} $.
$\mathcal{T}_{A\mu\nu}$ in Eq.(\ref{NYa}) is a nonzero torsion tensor that is used to identify the gravity, which generally depends on both the tetrad field and the spin connection,
\bqn
\mathcal{T}^{\lambda}_{\mu\nu} = 2 e^{\lambda}_a \left( \partial_{[\mu}e^a_{\nu]} + \omega^a_{b [\mu } e^b_{\nu]} \right).
\eqn
The  $\omega^a_{b \mu } $ is the spin connection, which generally has the form $\omega^a_{b \mu } = (\Lambda^{-1})^a_c \partial_{\mu} \Lambda^c_b$ and $\omega_{ab \mu } = -\omega_{ba \mu } $ with $\Lambda^a_b$ being the matrix elements of Lorentz transformation.
The form of $\mathcal{\widetilde{T}}^{A\mu\nu}$ in Eq.(\ref{NYa}) is $\mathcal{\widetilde{T}}^{A\mu\nu} = (1/2) \varepsilon^{\mu\nu \rho \sigma} \mathcal{T}^{A}{}_{\rho \sigma} $.

\subsection{ The parity-violating symmetric teleparallel gravity}
\renewcommand{\theequation}{2.4.\arabic{equation}} \setcounter{equation}{0}

STG is another constrained metric-affine theory. It is formulated in a spacetime endowed with zero curvature and zero torsion and attributes gravity to the non-metricity.
The parity-violating symmetric teleparallel gravity is constructed as an extension of the STGR model by introducing a parity-violating term.
The full action of gravity is\cite{Li:2021mdp}
\bqn\lb{PVSTG}
S = \int d^4x  \sqrt{-g} \left[ -\f{\hat{R}}{2} -c\phi\epsilon^{\mu\nu\rho\sigma}Q_{\mu\nu\alpha}Q_{\rho\sigma}{}^{\alpha}-\f{1}{2}\hat{\nabla}_{\mu}(Q^{\mu} - \bar{Q}^{\mu})  + \f{1}{2} \nabla_{\mu}\phi \nabla^{\mu}\phi - V(\phi) \right]  + S_m,
\eqn
where $\hat{R}$ is the curvature scalar, $c$ is a coupling constant, $\phi$ represents a scalar field, $S_m$ is other matter, $Q_{\alpha\mu\nu}$ is the non-metricity tensor and is defined as
\bqn
Q_{\alpha\mu\nu} \equiv \nabla_{\alpha} g_{\mu\nu} = \partial_{\alpha} g_{\mu\nu} - \Gamma^{\lambda}{}_{\alpha\mu}g_{\lambda\nu} - \Gamma^{\lambda}{}_{\alpha\nu}g_{\mu\lambda},
\eqn
the vectors $Q_{\mu},~\bar{Q}_{\mu}$ are two different traces of the non-metricity tensor and are given by
\bqn
Q_{\mu} = g^{\alpha\beta}Q_{\mu\alpha\beta},~~~\bar{Q}_{\mu} = g^{\alpha\beta}Q_{\alpha\beta\mu},
\eqn
$\hat{\nabla}$ denotes the covariant derivative associated with the Levi-Civit\'a connection.
It should be noted here that in the zero curvature condition, the affine connection can usually be expressed as
\bqn
\Gamma^{\lambda}{}_{\mu\nu} = (\Omega^{-1})^{\lambda}{}_{\sigma} \partial_{\mu}\Omega^{\sigma}{}_{\nu},
\eqn
where $\Omega^{\sigma}{}_{\nu}$ is an arbitrary element of the group $\rm{ GL(4) }$ and has non-zero determinant. The zero-torsion condition requires that $\Omega^{\sigma}{}_{\nu}$ should be expressed as $\Omega^{\sigma}{}_{\nu} = \partial y^{\sigma}/\partial x^{\nu}$, where $y^{\sigma}(x)$ are four functions that fully determine all components of the affine connection,
\bqn
\Gamma^{\lambda}{}_{\mu\nu} (x^{\mu})= \f{\partial x^{\lambda}}{\partial y^{\beta}} \partial_{\mu}\partial_{\nu}y^{\beta}.
\eqn
The above equation shows that the four functions $y^{\alpha}$ constitute a special coordinate system in which the affine connection vanishes, i.e. $ \Gamma^{\lambda}{}_{\mu\nu}= 0$.
Thus, the metric $g_{\mu\nu}$ and the function $y^{\alpha}(x)$ can be treated as independent variables in this theory.

\section{Applications of the parity-violating gravity to the isolated Sources \label{sec3}}
\renewcommand{\theequation}{3.\arabic{equation}} \setcounter{equation}{0}

In this section, based on the previous introduction to parity-violating gravity, we will present the study of gravitational wave propagation from isolated sources under these different frameworks.
We can assume that GWs are propagating on a homogeneous and isotropic background. In the case of a flat Friedmann-Robertson-Walker (FRW) universe, the spatial metric is written as
\bqn\lb{gij}
g_{ij}=a^2(\tau) (\delta_{ij} + h_{ij}(\tau, x^i)),
\eqn
where $\tau$ denotes the conformal time, which relates to the cosmic time $t$ by $dt=ad\tau$, and $a$ is the scale factor of the universe. Throughout this paper, we set the present scale factor $a_0=1$. $h_{ij}$ is the GW, which represents the transverse and traceless metric perturbations, i.e.,
\bqn\lb{hxwj}
\partial^i h_{ij} =0 = h_i^i.
\eqn

\subsection{ The ghost-free parity-violating gravity}
\renewcommand{\theequation}{3.1.\arabic{equation}} \setcounter{equation}{0}

As we have discussed in the previous section, CS gravity could be regarded as a specific version of ghost-free parity-violating gravity. The effect of the CS correction is included in the final result of the parity-violating corrections. Therefore the application of CS gravity is not discussed separately and is presented as a special case of ghost-free parity-violating gravity.

\subsubsection{ Gravitational waves in ghost-free parity-violating gravity }

With the above choice of background and the definition of GWs, the equation of motion of GWs can be derived. Substituting the metric perturbation into the action (\ref{xaction}) and expanding it to the second order in $h_{ij}$, the tensor quadratic action reads \cite{Qiao:2019wsh}
\bqn
S^{(2)} = \frac{1}{16\pi G} \int d\tau d^3 x a^4(\tau) \left[ \mathcal{L}_{\rm GR}^{(2)} + \mathcal{L}_{\rm PV}^{(2)}\right],
\eqn
where
\bqn
\mathcal{L}_{\rm GR}^{(2)} &=& \frac{1}{4 a^2} \left[ (h'_{ij})^2 - (\partial_k h_{ij})^2\right],\\
\mathcal{L}_{\rm PV}^{(2)} &=& \frac{1}{4 a^2} \left[\frac{c_1(\tau)}{a {M_{\rm PV}}} \epsilon^{ijk}h_{il}' \partial_j h_{kl}'+ \frac{c_2(\tau)}{a {M_{\rm PV}} }   \epsilon^{ijk}\partial^2h_{il} \partial_j h_{kl}\right].\nb\\
\eqn
Here, ${M_{\rm PV}}$ labels the parity-violating energy scale in this theory. $c_1$ and $c_2$ are the coefficients normalized by the energy scale ${M_{\rm PV}}$, which are given by Ref. \cite{Qiao:2019wsh}
\bqn
\frac{ c_1(\tau)}{{M_{\rm PV}}} &=& \dot{\vartheta}-4 \dot{a_1}\dot{\phi}^2 -8 a_1\dot{\phi}\ddot{\phi} + 8a_1 H \dot{\phi}^2- 2\dot{a_2}\dot{\phi}^2 - 4a_2\dot{\phi}\ddot{\phi}
-\dot{a_3}\dot{\phi}^2 - 2a_3\dot{\phi}\ddot{\phi} + 4a_3 H \dot{\phi}^2-8\dot{a_4}\dot{\phi}^2 -16a_4\dot{\phi}\ddot{\phi}\nb\\
&& -2 b_1\dot{\phi}^3 +4b_2\left( H\dot{\phi}^2-\dot{\phi}\ddot{\phi}\right)
 +2b_3  \dot{\phi}^3\ddot{\phi} + 2b_4 \dot{\phi}^3\ddot{\phi}
-2b_5 H \dot{\phi}^4+2b_7\dot{\phi}^3\ddot{\phi} + 6 b_7\dot{\phi}^4H,\label{c1-1}\\
\frac{c_2(\tau)}{{M_{\rm PV}}} &=&  \dot{\vartheta} - 2\dot{a_2}\dot{\phi}^2 -4a_2\dot{\phi}\ddot{\phi} +\dot{a_3}\dot{\phi}^2 +2a_3\dot{\phi}\ddot{\phi}
-8\dot{a_4}\dot{\phi}^2 -16a_4\dot{\phi}\ddot{\phi},\label{c2-2}
\eqn
{where a \textit{dot} denotes the derivative with respect to the cosmic time $t$, and $H \equiv \dot{a} /a$ is the Hubble parameter.
The ${M_{\rm PV}}$ can be constrained by solar system experiments and various astrophysical observations. As we have already described in the introduction, some of the parity-violating gravities have been examined by GW observations and give updated constraints on their respective corresponding energy scale ${M_{\rm PV}}$. For the ghost-free parity-violating gravity, using the results of the frame-dragging measurement with the GPB in the solar system experiment we have given the energy scale constraint: $M^{-1}_{\rm PV} \lesssim 10^4 \rm m$, which is consistent with the constraint obtained for CS gravity when the same factors are taken into account \cite{Qiao:2021fwi}. The latest constraint on $ M_{\rm PV}$ is also given using the event data from the GW detection to examine ghost-free parity-violating gravity, which we will describe later.}

We consider the GWs propagating in the vacuum, and ignore the source term. Varying the action with respect to $h_{ij}$, we obtain the field equation for $h_{ij}$,
\bqn
&&h_{ij}'' + 2 \mathcal{H} h_{ij}'  - \partial^2 h_{ij}
+ \frac{\epsilon^{ilk}}{a {M_{\rm PV}}} \partial_l \Big[ c_1 h_{jk}'' + (\mathcal{H}c_1+c_1') h_{jk}' - c_2 \partial^2 h_{jk}\Big]=0,
\eqn
where $\mathcal{H}\equiv a'/a$, and a {\emph{prime}} denotes the derivative with
respect to the conformal time $\tau$.

In the parity-violating gravities, it is convenient to decompose the GWs into circular polarization modes. To study the evolution of $h_{ij}$, we expand it over spatial Fourier harmonics,
\bqn\lb{fly}
h_{ij}(\tau, x^i) = \sum_{A={\rm R, L}} \int \frac{d^3 k}{(2\pi)^3}  h_A(\tau, k^i)e^{i k_i x^i} e_{ij}^{A}(k^i),
\eqn
where $e_{ij}^A$ denote the circular polarization tensors and satisfy the relation
\bqn
\epsilon^{i j k} n_i e_{kl}^A = i \rho_A e^{jA}_{l},
\eqn
with $\rho_{\rm R}=1$ and $\rho_{\rm L} =-1$. We find that the propagation equations of these two modes are decoupled, which can be cast into the form \cite{Qiao:2019wsh}
\bqn\lb{eom}
h_A'' + (2+\nu_A) \mathcal{H} h_A' + (1+\mu_A) k^2 h_A=0,
\eqn
where
\bqn
\nu_A &=& \frac{\rho_A k (c_1 \mathcal{H} -c_1' )/(a \mathcal{ H}  {M_{\rm PV}})}{1- \rho_A k c_1/(a {M_{\rm PV}})}, \lb{nuA}\\
\mu_A &=& \frac{\rho_A k (c_1 -c_2)/(a {M_{\rm PV}})}{1- \rho_A k c_1/(a {M_{\rm PV}})}.\lb{muA}
\eqn

\subsubsection{ Various effects of the parity violation }

In this subsubsection, we present the phase and amplitude corrections to the waveform of GWs arising from the parameters $\nu_A$ and $\mu_A$.
We further decompose $h_A$ as
\bqn
h_A = \bar h_A e^{-i \theta(\tau)}, \lb{decom} \\
\bar h_A = \mathcal{A}_A e^{- i \Phi(\tau)},\lb{decom1}
\eqn
where $\bar h_A$ satisfies
\bqn\lb{hAbar}
\bar h_A '' + 2 \mathcal{H} \bar h'_A + (1+\mu_A) k^2\bar h_A=0,
\eqn
 here $\mathcal{A}_A$ denotes the amplitude of $\bar h_A$ and $\Phi(\tau)$ is the phase. With the above decomposition,  $\theta (\tau)$ denotes the correction arising from $\nu_A$, while the corrections due to $\mu_A$ is included in $\bar h_A$.
Such processing allows the phase correction to be calculated separately before the amplitude correction is investigated.
Details of the calculation of the phase and amplitude corrections are given in the following as Ref. \cite{Qiao:2019wsh}.

\subsubsection{Phase modifications}

We first concentrate on the corrections arising from the parameter $\mu_A$, which leads to the velocity difference of the two circular polarizations of GWs. To proceed, we define $\bar u_k^{A}(\tau) = \frac{1}{2} a(\tau) M_{\rm Pl}\bar{h}_A (\tau) $ and then Eq. (\ref{hAbar}) can be written as
\bqn
\frac{d^2 \bar u_k^{A}}{d\tau^2} + \left(\omega_A^2 - \frac{a''}{a}\right) \bar u_k^{A}=0,
\eqn
where
\bqn
\omega_A^2(\tau) = k^2(1+\mu_A),
\eqn
is the modified dispersion relation. With this relation, the speed of the graviton reads
\bqn
{v_A^2} = {k^2}/{\omega_A^2} \simeq  1- \rho_A  (c_1-c_2) \left( {k}/{a {M_{\rm PV}}}\right),
\eqn
which leads to
\bqn
{v_A} \simeq 1- ({1}/{2}) \rho_A  (c_1-c_2) \left( {k}/{a {M_{\rm PV}}}\right).
\eqn
Since $\rho_A$ have the opposite signs for left-hand and right-hand polarization modes, we find that one mode is superluminal and the other is subluminal. Considering a graviton emitted radially at $r=r_e$ and received at $r=0$, we have
\bqn
\frac{dr }{dt} = - \frac{1}{a} \left[1- \frac{1}{2}\rho_A  (c_1-c_2) \left( \frac{k}{a {M_{\rm PV}}}\right)\right].
\eqn
Integrating this equation from the emission time ($r=r_e$) to arrival time ($r=0$), one obtains
\bqn
r_e = \int^{t_0}_{t_e} \frac{dt}{a(t)} - \frac{1}{2}\rho_A \left(\frac{k}{ {M_{\rm PV}}}\right) \int_{t_e}^{t_0} \frac{c_1-c_2}{a^{2}}dt.
\eqn

Consider gravitons with the same $\rho_{A}$ emitted at two different times $t_e$ and $t_e'$, with wave numbers $k$ and $k'$, and received at corresponding arrival times $t_0$ and $t_0'$ ($r_e$ is the same for both). Assuming $\Delta t_e\equiv t_e-t_e'\ll a/\dot{a}$, then the difference of their arrival times is given by
\[
\Delta t_{0}=(1+z)\Delta t_e+\frac{1}{2}\rho_A\frac{k-k'}{{M_{\rm PV}}}\int_{t_e}^{t_0} \frac{c_1-c_2} {a^{2}}dt,
\]
where $z\equiv 1/a(t_e)-1$ is the cosmological redshift.

Let us focus on the GW signal generated by non-spinning, quasi-circular inspiral in the post-Newtonian approximation. Relative to the GW in GR, the term $\mu_{\rm A}$ modifies the phase of GW $\Phi(\tau)$. The Fourier transform of $\bar{h}_{\rm A}$ can be obtained analytically in the stationary phase approximation, where we assume that the phase is changing much more rapidly than the amplitude, which is given by \cite{spa}
\bqn\lb{wtfly}
\tilde{\bar{h}}_{ A}(f)=\frac{{\mathcal{A}}_{ A}(f)}{\sqrt{df/dt}}e^{i\Psi_A(f)},
\eqn
where $f$ is the GW frequency at the detector, and $\Psi$ is the phase of GWs. In \cite{Mirshekari:2011yq}, it is proved that, the difference of arrival times as above induces the modification of GWs phases $\Psi_A$ as follows,
\bqn
\Psi_A(f) = \Psi_A^{\rm GR} (f) + \delta \Psi_A(f),
\eqn
with
\bqn \lb{delta_Phi}
\delta \Psi_A(f) = \xi_A u^{2},
\eqn
where
\bqn
\xi_A &=& \frac{\rho_A}{{M_{\rm PV}} \mathcal{M}^{2}} \int_{t_e}^{t_0} {\frac{c_1-c_2}{a^{2}}}dt, \\
u&=& \pi \mathcal{M} f.
\eqn
The quantity $\mathcal{M} = (1+z) \mathcal{M}_{\rm c}$ is the measured chirp mass, and $\mathcal{M}_{\rm c}\equiv (m_1 m_2)^{3/5}/(m_1+m_2)^{1/5}$ is the chirp mass of the binary system with component masses $m_1$ and $m_2$.

\subsubsection{Amplitude modifications}

Now, let us turn to study the effect caused by $\nu_A$. Plugging the decomposition (\ref{decom1}) into (\ref{hAbar}), one finds the equation for $\Phi(t)$,
\bqn\lb{eom_Phi}
i \Phi'' + \Phi'^2 + 2 i \mathcal{H} \Phi' - (1+\mu_A)k^2=0.
\eqn
Similarly, plugging the decomposition (\ref{decom}) and (\ref{decom1}) into (\ref{eom}), one obtains
\bqn \lb{eom2}
&&i (\theta''+\Phi'') + (\Phi'+\theta')^2
+i (2+\nu_A)\mathcal{H} (\theta'+\Phi') - (1+\mu_A)k^2=0.
\eqn
Using the equation of motion (\ref{eom_Phi}) for $\Phi$, the above equation reduces to
\bqn
i \theta''+ 2 \theta' \Phi' + \theta'^2 + i (2+\nu_A)\mathcal{H} \theta'+ i \nu_A \mathcal{H} \Phi'=0.\nb\\
\eqn
The phase $\Phi$ is expected to be close to that in GR $\Phi_{\rm GR}$, and $\Phi_{\rm GR}'  \sim k$, where the wave number relates to the GW frequency by $k=2\pi f$. Assuming that
\bqn
\theta'' \ll \Phi'\theta'  \sim k \theta',\;\; k \gg \mathcal{H},
\eqn
and keeping only the leading-order terms, the above equation can be simplified into the form
\bqn
2  \theta' + i \mathcal{H} \nu_A  =0,
\eqn
which has the solution
\bqn
 \theta = -\frac{i}{2} \int_{\tau_e}^{\tau_0} \mathcal{H} \nu_A d\tau.
 \eqn
We observe that the contribution of $\nu_A$ in the phase is purely imaginary. This indicates that the parameter $\nu_A$ leads to modifications of the amplitude of the GWs during the propagation. As a result, relative to the corresponding mode in GR, the amplitude of the left-hand circular polarization of GWs will increase (or decrease) during the propagation, while the amplitude for the right-hand mode will decrease (or increase).

More specifically, one can write the waveform of GWs with parity violation effects in the form
\bqn\lb{waveformA}
h_A = h_{A}^{\rm GR} (1+\delta h_A) e^{ - i \delta \Phi_A},
\eqn
where
\bqn\lb{waveform}
1+ \delta h_A &=&  \exp\left(-\frac{1}{2} \int_{\tau_e}^{\tau_0} \mathcal{H} \nu_A d\tau \right),
\eqn
and $\delta \Phi_A$ is given by (\ref{delta_Phi}). Noticing that
\bqn
\frac{1}{2} \nu_A \mathcal{H} = \frac{1}{2} \left[\ln\left(1- \rho_A \frac{k c_1 }{a {M_{\rm PV}}}\right)\right]',
\eqn
we find
\bqn
1+\delta h_A &=& \sqrt{ \frac{1- \rho_A  k c_1(\tau_e) / [a (\tau_e) {M_{\rm PV}}] }{1- \rho_A  k c_1(\tau_0) / [a (\tau_0) {M_{\rm PV}}] } }\nb\\
&\simeq & 1 + \frac{1}{2} \rho_A k \left(\frac{c_1(\tau_0)}{a(\tau_0) {M_{\rm PV}}} - \frac{c_1(\tau_e)}{a(\tau_e) {M_{\rm PV}}}\right),
\eqn
which gives
\bqn
\delta h_A & \simeq &  \frac{1}{2} \rho_A k \left(\frac{c_1(\tau_0)}{a(\tau_0) {M_{\rm PV}}} - \frac{c_1(\tau_e)}{a(\tau_e) {M_{\rm PV}}}\right)\nb\\
&=& \rho_A \frac{ \pi f}{ {M_{\rm PV}}} \Big[c_1(\tau_0) - (1+z) c_1(\tau_e)\Big].
\eqn
Using $u$ and $\mathcal{M}$, one can rewrite $\delta h_A$ in the form
\bqn\label{delta_h}
\delta h_A = \frac{\rho_{A} u}{ {M_{\rm PV}}\mathcal{M}}  \Big[c_1(\tau_0) - (1+z) c_1(\tau_e)\Big].
\eqn
This relation indicates that the amplitude birefringence of GWs depends only on the values of the coefficient $c_1$ at the emitting and observed times.

\subsubsection{Post-Newtonian orders of the correction terms}

In general, we can write the GWs in the Fourier domain. Similar to the parameterized post-Einsteinian framework of GWs developed in \cite{Yunes:2009ke, Cornish:2011ys}, for each circular polarization mode, we can also write the GW waveform as the following parameterized form
\bqn\label{hA-f}
\tilde{h}_A(f) = \tilde{h}_{A}^{\rm GR} (1+ \alpha_{A}^{\rm ppe} u^{a_{A}^{\rm ppe}})e^{i \beta_{A}^{\rm ppe} u^{b_{A}^{\rm ppe}}},
\eqn
where $\alpha_{A}^{\rm ppe} u^{a_{A}^{\rm ppe}}=\delta h_A$ and $\beta_{A}^{\rm ppe} u^{b_{A}^{\rm ppe}}=\delta\Psi_{A}$ represent the amplitude and phase modification respectively. These two terms capture non-GR modifications in the waveform in a generic way. The coefficients $a_A^{\rm ppe}$ and $b_A^{\rm ppe}$ indicate the post-Newtonian (PN) orders of these modifications. In comparison with the results derived in the previous subsection, we obtain that
\bqn
\alpha_{A}^{\rm ppe} &=& \frac{\rho_{A}}{ {M_{\rm PV}}\mathcal{M}}  \Big[c_1(\tau_0) - (1+z) c_1(\tau_e)\Big], \\
a_{A}^{\rm ppe} &=& 1, \\
\beta_{A}^{\rm ppe} &=&  \xi_A,  \\
b_{A}^{\rm ppe} &=& 2.
\eqn
Let us now count the PN order of these parity-violating corrections relative to GR. The relative correction from GR is said to be $n$ PN order if it is proportional to $f^{2n/3}$. Thus, the amplitude correction enters at 1.5 PN order, and the phase correction enters at 5.5 PN order (note that the phase of GR $\propto f^{-5/3}$ at leading order).

In order to make contact with observations, it is convenient to analyze the GWs in the Fourier domain, and the responses of detectors for the GW signals $\tilde{h}(f)$ can be written in terms of the waveform of $\tilde{h}_+$ and $\tilde{h}_\times$ as
\bqn
\tilde{h}(f) = [F_+ \tilde{h}_+(f) + F_\times \tilde{h}_\times(f)] e^{- 2 \pi i f \Delta t},
\eqn
where $F_{+}$ and $F_{\times}$ are the beam pattern functions of GW detectors, depending on the source location and polarization angle \cite{Jaranowski:1998qm,Zhao:2017cbb}. $\Delta t$ is the arrival time difference between the detector and the geocenter. In GR, the waveform of the two polarizations $\tilde{h}_+(f)$ and $\tilde{h}_\times (f)$ are given by
\bqn
\tilde{h}^{\rm GR}_+ = (1+ {\chi^2}) \mathcal{A}e^{i \Psi}, \;\; \\
\tilde{h}^{\rm GR}_\times = 2  {\chi} \mathcal{A} e^{i (\Psi+\pi/2)},
\eqn
where $\mathcal{A}$ and $\Psi$ denote the amplitude and phase of the waveforms $h^{\rm GR}_{+ \; \times}$, and $\chi=\cos\iota$ with $\iota$ being the inclination angle of the binary system.
Now we would like to derive how the parity violation can affect both the amplitude and the phase of the above waveforms. The circular polarization modes $\tilde{h}_{R}$ and $\tilde{h}_L$ relate to the modes $\tilde{h}_+$ and $\tilde{h}_{\times}$ via
\bqn
\tilde{h}_+ =  \frac{\tilde{h}_{L} + \tilde{h}_{R}}{\sqrt{2}},~~
\tilde{h}_\times =  \frac{\tilde{h}_{L} - \tilde{h}_{R}}{\sqrt{2} i}.
\eqn
The Fourier waveform $\tilde{h}(f)$ becomes \cite{Qiao:2019wsh}
\bqn\label{final-hf}
\tilde{h}(f)= \mathcal{A} \delta \mathcal{A} e^{i (\Psi +\delta \Psi )} ,
\eqn
where
\bqn\label{final-delta}
\delta \mathcal{A} &=& \sqrt{(1+ {\chi^2})^2F^2_+ +4 {\chi}^2F^2_\times}
\times\Big[1+\frac{2 {\chi}(1+ {\chi}^2)(F^2_+ + F^2_\times)}{(1+ {\chi}^2)^2F^2_+ +4 {\chi}^2F^2_\times}\delta h
-\frac{(1- {\chi}^2)^2F_+F_\times}{(1+ {\chi}^2)^2F^2_+ +4 {\chi}^2F^2_\times}\delta\phi+\mathcal{O}((\delta h)^2,(\delta\phi)^2)\Big] ,\nb\\
\delta \Psi
&=&\tan^{-1}\left[\frac{2  {\chi} F_\times}{(1+ {\chi}^2)F_+}\right]+\frac{(1- {\chi}^2)^2 F_+ F_\times}{(1+ {\chi}^2)^2 F^2_+ + 4  {\chi}^2F^2_\times}\delta h
+\frac{2  {\chi} (1+  {\chi}^2)(F^2_+ + F^2_\times)}{(1+  {\chi}^2)^2 F^2_+ + 4  {\chi}^2F^2_\times}\delta\phi+\mathcal{O}((\delta h)^2,(\delta\phi)^2),
\eqn
with $\delta\phi$ and $\delta h$ corresponding to Eq. (\ref{delta_Phi}) and Eq. (\ref{delta_h}) respectively.

\subsubsection{ Observational properties and constraints }
The final expression for the waveform clearly shows that the modifications to the GWs relative to the waveform in GR are quantified by the terms $\delta h$ and $\delta \phi$, both of which are induced by the parity-violating terms. $\delta h$ and $\delta \phi$ are amplitude birefringence and velocity birefringence effects, respectively, between the left- and right-hand polarization modes. In the specific case with $\delta h=\delta\phi=0$, the formula in (\ref{final-hf}) returns to that in GR. In the CS modified gravity with $\delta \phi=0$ and $\delta h\neq 0$, the formulas in Eq.(\ref{final-delta}) returns to the corresponding ones in \cite{Yagi:2017zhb}. Eq. (\ref{final-hf}) also indicates that the evolution of polarization modes ${h}_+$ and $h_{\times}$ are not independent, the mixture of two modes is inevitable. This explains the presence of terms $\delta h$ and $\delta\phi$  that appear in the phase and amplitude modifications of $\tilde{h}(f)$.
A further extension to the investigation of corrections to GWs waveforms is presented in Ref. \cite{Zhao:2019xmm}, where the correction terms contain both parity-conserving and parity-violating terms.
The amplitude and phase corrections induced by the parity-conserving terms modified only the amplitude and phase of $\tilde{h}(f)$, respectively, i.e. the parity-conserving terms affect the evolution of the two polarization modes independently.

In the leading order, the modification $\delta\mathcal{A}$ (or $\delta\Psi$) linearly depends on $\delta h$ and $\delta \phi$. Estimating their relative magnitudes is valuable for examining waveform corrections.
Assuming that the GW is emitted at redshift $z\sim O(1)$ and approximately treating $c_1$ and $c_2$ as constants in the GW propagation process, one obtains the ratio of the two correction terms as $\delta\phi/\delta h \sim t_0 f$, where $f$ is the GW frequency and $t_0=13.8$ billion years is the age of the Universe. As known, $f\sim 100$ Hz for the ground-based GW detectors, and $f\sim 0.01$ Hz for the space-borne detectors. In both cases, the results suggest that $\delta\phi$ is more than ten orders of magnitude larger than $\delta h$. Therefore, we arrive at the conclusion: In the general ghost-free parity-violating gravities, both the amplitude and phase corrections of GW waveform $\tilde{h}(f)$ predominantly come from the contribution of velocity birefringence rather than that of the amplitude birefringence.

In order to seek birefringence effects in waveforms produced by parity violation, Ref. \cite{Wang:2021gqm} direct comparison with GW data was performed using Bayesian inference. For all GW events, there was no indication of parity violations found in the results. Meanwhile Ref. \cite{Wang:2021gqm} has given the $90\%$ lower limit for MPV is $0.09 \rm GeV$, which is the tightest constraint on $M_{\rm PV}$ up-to-date. The velocity birefringence effect is stronger for waveform correction than the amplitude birefringence effect, and  Ref. \cite{Wang:2021gqm} also gave a constraint of $M_{\rm PV} > 1 \times 10^{-22} \rm GeV$ by considering only the amplitude birefringence correction, which is consistent with CS gravity.

\subsection{ The parity-violating Ho\v{r}ava-Lifshitz gravity}
\renewcommand{\theequation}{3.2.\arabic{equation}} \setcounter{equation}{0}

\subsubsection{ Gravitational waves in the parity-violating Ho\v{r}ava-Lifshitz gravity }
The general formulas of the linearized tensor perturbations were given in \cite{Zhu:2013fja}, so in the rest of this section we give a very brief summary of the main results obtained there. Consider a flat FRW universe and assuming that matter fields have no contributions to tensor perturbations, the quadratic part of the total action can be cast in the form \cite{Zhao:2019xmm},

\bqn
S^{(2)} &=& \f{1}{16\pi G} \int d\tau d^3x\Bigg[ \f{a^2}{4}(h'_{ij})^2- \f{1}{4}a^2(\partial_kh_{ij})^2 - \f{\hat{\gamma}_3}{4M^2_{\rm PV}}(\partial^2h_{ij})^2 - \f{\hat{\gamma}_5}{4M^4_{\rm PV} a^2}  (\partial^2\partial_k h_{ij})^2 \nb\\
          &&-\f{\alpha_1 a \epsilon^{ijk}}{2M_{\rm PV}}(\partial_l h^m_i \partial_m \partial_j h^l_k - \partial_l h_{im} \partial^l \partial_j h^m_k)  - \f{\alpha_2\epsilon^{ijk}}{4M^3_{\rm PV} a} \partial^2 h_{il}   (\partial^2h^l_k)_{,j}
          - \f{3\alpha_0\mathcal{H}}{8M_{\rm PV}a}(\partial_k h_{ij})^2 \Bigg],
\eqn
where $\hat{\gamma}_3 \equiv (2M_{\rm PV}/M_{\rm Pl})^2 \gamma_3$ and  $\hat{\gamma}_5 \equiv (2M_{\rm PV}/M_{\rm Pl})^4 \gamma_5$, and $ \gamma_3$ and $ \gamma_5$ are the dimensionless coupling constants of the theory.
To avoid fine-tuning, $\alpha_n$ and $\hat{\gamma}_n$ are expected to be of the same order. Following the variational principle, the equations of motion for $h_{ij}$ read,
\bqn
h''_{ij} + 2\mathcal{H} h'_{ij} - \alpha^2\partial^2 h_{ij} + \f{\hat{\gamma}_3}{a^2M^2_{\rm PV}} \partial^4h_{ij} - \f{\hat{\gamma}_5}{a^4M^4_{\rm PV}} \partial^6h_{ij} + \epsilon_i{}^{lk}\left( \f{2\alpha_2}{aM_{\rm PV}} + \f{\alpha_2}{a^3M^3_{\rm PV}} \partial^2\right) (\partial^2h_{jk})_{,l} = 0,
\eqn
where $\alpha^2 \equiv 1+ 3\alpha_0\mathcal{H}/(2aM_{\rm PV})$. To study the evolution of $h_{ij}$, we expand it over spatial Fourier harmonics. For each circular polarization mode, the standard parameterization of the equation of motion for GW takes the form \cite{Zhao:2019xmm}
\bqn\lb{HLeq}
h_A'' + (2+\nu_A + \bar{\nu}) \mathcal{H} h_A' + (1+\mu_A + \bar{\mu}) \alpha^2 k^2 h_A=0,
\eqn
where
\bqn
\bar{\nu} =0 = \nu_A,~~\bar{\mu} = \delta_2(k/a M_{\rm PV})^2 + \delta_4(k/a M_{\rm PV})^4 + 3\alpha_0\mathcal{H}/(2M_{\rm PV}a),~~\mu_A = \delta_1 \rho_A (k/a M_{\rm PV}) - \delta_3\rho_A(k/a M_{\rm PV})^3,\nb\\
\eqn
with $\delta_1 \equiv 2\alpha_1/\alpha^3,~\delta_2 \equiv \hat{\gamma}_3/\alpha^4,~\delta_3 \equiv \alpha_2/\alpha^5,~\delta_4 \equiv \hat{\gamma}_5/\alpha^6$. In the late universe, $a \sim 1$, and $\mathcal{H} \ll M_{\rm PV}$, so we find $\alpha^2 \rightarrow 1$. In the expression of $\bar{\mu}$, the second term is always negligible, and the relative magnitude of the first and third terms depends on the values of $k$ and $M_{\rm PV}$. In the theory, which includes both the third- and fifth-order operators, the first term in $\mu_A$ is dominant. While for the theory, which includes only the fifth-order operator, only the second term in $\mu_A$ exists.

\subsubsection{ Observational properties  }

Eq. (\ref{HLeq}) reflects the fact that both the parity-conserving and parity-violating terms are corrected for the dispersion relation $\omega_A =(1+\mu_A + \bar{\mu}) \alpha^2 k^2$, which is portrayed by the parameters $\bar{\mu}$ and $\mu_A$ respectively.
 It can be clearly seen that in the case of $\mu_A=0$, although the dispersion relation is corrected, the propagation velocities of the two circularly polarised modes are the same. Only when the dispersion relation is corrected by the parity-violating terms $\mu_A$, the propagation velocities of the two circular polarization modes are handedness-dependent, thus producing a velocity birefringence effect. This suggests that the birefringence effect is a powerful support for testing the existence of parity violation in gravity.

\subsection{ The Nieh-Yan modified teleparallel gravity }
\renewcommand{\theequation}{3.3.\arabic{equation}} \setcounter{equation}{0}

In section II, we have given a brief introduction to the modified teleparallel gravity. Here, we will present a discussion of the study of GWs propagation in this gravity \cite{Wu:2021ndf}.
Following the variational principle, the equation of motion is obtained by variation of the action (\ref{NYa}) as
\bqn
G^{\mu\nu} + N^{\mu\nu} = T^{\mu\nu} + T^{\mu\nu}_{\theta},
\eqn
where $G^{\mu\nu}$ is the Einstein tensor, $T^{\mu\nu} $ and $T^{\mu\nu}_{\theta}$ are the energy-momentum tensor for the matter and the scalar field respectively,
and the tensor $N^{\mu\nu}$ is of the form $N^{\mu\nu} = c e^{\nu}{}_{A} \partial_{\rho} \theta \mathcal{\widetilde{T}}^{A\mu\rho} = c \partial_{\rho} \theta \mathcal{\widetilde{T}}^{\nu\mu\rho}$.
The antisymmetric part of the tensor $N^{\mu\nu}$ is vanishing, which means $N^{\mu\nu}$ is subject to a symmetry constraint as
\bqn\lb{smny}
N^{[\mu\nu]} = 0.
\eqn
The variation of the action (\ref{NYa}) with respect to the scalar field $\theta$ obtains the equation of motion as
\bqn
b\square\theta + bV_{\theta} - \f{c}{4} \mathcal{{T}}_{A\nu\mu} \mathcal{\widetilde{T}}^{A\nu\mu}=0,
\eqn
where $V_{\theta}$ denotes the first derivative of the potential to the scalar field.
It is interesting to mention here that the value of the parameter $b$ determines the different versions of this gravity. Similar to CS gravity, when $b = 0$ or $b\neq 0$ correspond to the non-dynamic or dynamic versions respectively.
As shown in \cite{{Li:2020xjt,Li:2021wij}}, the propagation of GW in both theories follows the same propagation equation, therefore in the following, we will not distinguish between the two versions and take b = 1.

\subsubsection{ Gravitational waves in Nieh-Yan modified teleparallel gravity }

We have given the perturbation of the spatial metric in Eq. (\ref{gij}), the tetrad field perturbation is \cite{Wu:2021ndf}
\bqn
e^0_0=a,~e^0_i=0,~e^a_0=0, \nb\\
e^a_i = a\left( \gamma^a_i + \f{1}{2}\gamma^{aj}h_{ij} \right),
\eqn
where $\gamma^a_i$ can be regarded as a space tetrad on a three-dimensional space hypersurface. For a flat universe, there is the relation $\delta_{ij} = \delta_{ab} \gamma^a_i \gamma^b_j$. It is important to note here that the tensor perturbation comes only from the tetrad field, and the spin connections or local Lorentzian matrices do not contribute to the tensor perturbation.

The above tetrad field and metric perturbation are substituted into the action (\ref{NYa}) and extended to the second order in $h_{ij}$. After tedious calculations, the tensor perturbation form of the action is
\bqn\lb{zleca}
S^{(2)} =\frac{1}{8\pi G}  \int d^4x \f{a^2}{8} (h'_{ij}h'_{ij} - \partial_k h_{ij} \partial^k h^{ij} - c\theta' \epsilon_{i j k} h_{il}\partial_jh_{kl}).
\eqn
To facilitate the study of the physics, GWs are usually assumed to propagate in a vacuum, ignoring the source term. Varying the action with respect to $h_{ij}$, the equations of propagation are
\bqn
&&h_{ij}'' + 2 \mathcal{H} h_{ij}'  - \partial^2 h_{ij}
+ \frac{1}{2} c\theta'  \left( \epsilon_{lki} \partial_l h_{jk} +  \epsilon_{lkj} \partial_l h_{ik}  \right)=0.
\eqn
We again substitute the propagation equation using the expanded form of $h_{ij}$ (\ref{fly}) over the spatial Fourier harmonics. The propagation equation for the two modes becomes the standard parameterized form(\ref{eom})
\bqn\lb{nyeom}
h_A'' + (2+\nu_A) \mathcal{H} h_A' + (1+\mu_A) k^2 h_A=0,
\eqn
where
\bqn
\nu_{A} = 0, ~~~\mu_{A} = \f{\rho_A c\theta'}{k}.
\eqn
This equation shows that left- and right-handed polarized GWs propagate with different velocities.

\subsubsection{ The effects of the parity violation }

It can be observed from Eq.(\ref{nyeom}) that the significant effect of the violating-parity term on GW is the correction of the dispersion relation $\omega_A = k^2 \left(1+ \rho_A\f{c\theta'}{k}\right)$.
Considering small coupling constant $c$ and slow evolution of $\theta$, one can find from the dispersion relation that GWs with different helicities have different phase velocities
\bqn
v_A \simeq 1+  \rho_A\f{c\theta'}{2k} = 1+  \rho_A\f{aM_{\rm PV}}{2k},
\eqn
where $M_{\rm PV} = c\theta'/a = c\dot{\theta}$ is a characteristic energy scale.
Considering a graviton emitted radially at $r=r_e$ and received at $r=0$, we have
\bqn
\frac{dr }{dt} = - \frac{1}{a} \left[1+  \rho_A\f{aM_{\rm PV}}{2k}  \right].
\eqn
Integrating this equation from the emission time ($r=r_e$) to arrival time ($r=0$), one obtains
\bqn
r_e = \int^{t_0}_{t_e} \frac{dt}{a(t)} +  \rho_A \frac{{M_{\rm PV}} }{ 2k } \int_{t_e}^{t_0}dt.
\eqn

Consider gravitons with the same $\rho_{A}$ emitted at two different times $t_e$ and $t_e'$, with wave numbers $k$ and $k'$, and received at corresponding arrival times $t_0$ and $t_0'$ ($r_e$ is the same for both). Assuming $\Delta t_e\equiv t_e-t_e'\ll a/\dot{a}$, then the difference of their arrival times is given by
\[
\Delta t_{0}=(1+z)\Delta t_e+\frac{\rho_A}{2}\left( \frac{M_{\rm PV}}{k'}- \frac{M_{\rm PV}}{k} \right)\int_{t_e}^{t_0} dt,
\]
where $z\equiv 1/a(t_e)-1$ is the cosmological redshift.

Therefore, the parity violation due to the Nieh-Yan term changes the phase of the GWs relative to the GWs in the GR. The expression $h_A$(\ref{wtfly}), computed analytically from the stationary phase approximation in the Fourier domain, could be directly applied. The correction of the GWs phase $\Psi$ due to different arrival times is as follows,
\bqn
\Psi_A(f) &=& \Psi^{GR}_A(f) + \rho_A\delta\Psi_1(f),
\eqn
where
\bqn
\delta\Psi_1(f) &=& A_{\mu} \ln u\nb\\
A_{\mu} &= & \f{M_{PV}}{2H_0} \int^z_)\f{dz}{(1+z)\sqrt{(1+z)^3\Omega_m + \Omega_{\Lambda}}}.
\eqn
Combining this modified phase $\Psi$ and the relationship between $h_{+,\times}$ and $h_{\rm R,L}$, the Fourier waveform $h(f)$ becomes
\bqn
h(f) = \mathcal{A} \delta \mathcal{A} e^{i(\Psi + \delta\psi)},
\eqn
where
\bqn
\delta \mathcal{A}&=&\sqrt{(1+ {\chi^2})^2F^2_+ +4 {\chi}^2F^2_\times}
\times\Big[1-\frac{(1- {\chi}^2)^2F_+F_\times}{(1+ {\chi}^2)^2F^2_+ +4 {\chi}^2F^2_\times}\delta\Psi_1\Big] ,\nb\\
\delta \Psi
&=&\tan^{-1}\left[\frac{2  {\chi} F_\times}{(1+ {\chi}^2)F_+}\right]
+\frac{2  {\chi} (1+  {\chi}^2)(F^2_+ + F^2_\times)}{(1+  {\chi}^2)^2 F^2_+ + 4  {\chi}^2F^2_\times}\delta\Psi_1.
\eqn

\subsubsection{ Observational properties and constraints }

The expression for this waveform represents that the Nieh-Yan term produces only a velocity birefringence effect, which is the opposite of CS theory. In contrast to parity-violating scalar-tenser gravity, the Nieh-Yan term does not produce an amplitude birefringence effect. It is consistent with our expectation that the Nieh-Yan term in the equation of motion (\ref{nyeom}) is only modified by the dispersion relation. To further investigate the correction of the waveform by velocity birefringence, Ref.\cite{Wu:2021ndf} performs full Bayesian inference on the 46 GW events of BBH in the LIGO-Virgo catalogs GWTC-1 and GWTC-2. The results revealed no indication of any parity violation due to the parity-violating Nieh-Yan term, and placed an upper bound on the energy scale $M_{\rm PV}$ of $M_{\rm PV} < 6.5 \times 10^{-42} \rm GeV$ with a confidence level of $90\%$. It represents the first constraint so far on the Nieh-Yan correction for teleparallel gravity.

\subsection{ The parity-violating symmetric teleparallel gravity}
\renewcommand{\theequation}{3.4.\arabic{equation}} \setcounter{equation}{0}

\subsubsection{ Gravitational waves in parity-violating symmetric teleparallel gravity }

Following the parity-violating gravities presented in section II, here we will analyze the effect of the parity-violating term on the propagation of GWs in the parity-violating symmetric teleparallel gravity.
After tedious calculations in Ref. \cite{Li:2021mdp}, the perturbation form of the action (\ref{PVSTG}) is
\bqn\lb{pvstmg}
S^{(2)} =\frac{1}{8\pi G} \int d^4x a^2 \left[\f{1}{8}(h'_{ij}h'_{ij} - \partial_k h_{ij} \partial^k h^{ij} )- 2c(2\mathcal{H}\phi + \phi') \epsilon_{i j k} h_{il}\partial_jh_{kl} \right].
\eqn
Considering GWs propagation in a vacuum and ignoring the source term, the equation of motion of GWs is
\bqn
&&h_{ij}'' + 2 \mathcal{H} h_{ij}'  - \partial^2 h_{ij}
-4 c(2\mathcal{H}\phi + \phi') \left( \epsilon_{lki} \partial_l h_{jk} +  \epsilon_{lkj} \partial_l h_{ik}  \right)=0.
\eqn
By replacing the propagation equations with the extended form of $h_{ij}$ on the spatial Fourier harmonics, the propagation equations for the two modes become the standard parametric form as
\bqn\lb{pvstg}
h_A'' + (2+\nu_A) \mathcal{H} h_A' + (1+\mu_A) k^2 h_A=0,
\eqn
where
\bqn
\nu_{A} = 0, ~~~\mu_{A} =  \f{a \rho_AM _{\rm PV} }{k},
\eqn
with
\bqn
M _{\rm PV} \equiv -8 \f{c( 2\mathcal{H}\phi + \phi')}{a} .
\eqn
{This equation of motion shows that the parity violation term only changes the dispersion relation in the propagation equation, which is similar to Nieh-Yan modified teleparallel gravity.}

\subsubsection{ Propagation speed and constraints }

Meanwhile, Eq.(\ref{pvstg}) reflects that the significant effect of the parity-violating term on the GW is to induce a handedness-dependent dispersion relation $\omega^2 = (1+ \mu_A)k^2= k^2 + a \rho_A M _{\rm PV} k $, which in turn leads to a difference between the propagation velocities of the two helicities of the GW. This phenomenon, also known as the velocity birefringence of GW, characterizes the parity violation of the theory.
Assuming again that the small coupling constants $c$ and slow evolution of $\phi$, it can be found from the dispersion relation that GWs with different helicities have different phase velocities as
\bqn
v^A_p = \f{\omega_A}{k} \approx 1+ \f{\rho_A a M _{\rm PV} }{2k}.
\eqn
These propagating velocities of GWs differ from the speed of light. This deviation is tightly constrained by current GWs experiments.
Ref. \cite{Wu:2021ndf} has targeted this velocity birefringence correction to the GW waveform by using LIGO-Virgo observations of event data from GWs of binary black hole mergers with a tighter constraint on $M_{\rm PV}: M _{\rm PV} < 6.5 \times 10^{-42} \rm GeV$.

\subsection{Intercomparison of parity-violating gravities}

In the previous subsections, we have described the effect of different parity-violating gravities on the propagation of gravitational waves produced by isolated sources. The results for the ghost-free parity-violating gravity show both velocity birefringence and amplitude birefringence effects from the parity-violating terms, while the other three gravities produce only the velocity birefringence effect. These birefringence effects lead to different corrections to the waveforms of GWs. In particular, the corrections to the waveform come mainly from the velocity birefringence effect compared to the amplitude birefringence effect in the ghost-free parity-violating gravity. Based on these corrected waveforms, the parity-violating gravities are examined separately and analyzed using a selection of currently observed GWs events. The results present that there is no significant parity violation in parity-violating gravities. In addition, the parity-violating gravities are given separately with the latest constraints on the energy scale: $M_{\rm PV} > 0.09 \rm GeV$ for the ghost-free parity-violating gravity and $M_{\rm PV}<6.5 \times 10^{-42} \rm GeV$ for the Nieh-Yan modified teleparallel gravity and the parity-violating symmetric teleparallel gravity.


We should mention that, it is a systematic issue to prove or falsify general relativity or other gravitational theory through gravitational-wave observation. The different properties (including the Lorentz symmetry, parity symmetry, equivalence principle, velocity, polarization, mass and dispersion relation of GWs, the possible evolution of gravitational constant $G$ and so on) of theory need to be tested at the same time with GWs at different frequency ranges. Actually, in addition to the birefringence effects in the general PV gravities as discussed above, some other characters of gravitational waves might exist in the general modified gravities \cite{brans-dicke,gongyg,smg,einstein-aether,Li:2022grj}. For instance, the extra polarization modes, except for the ``plus" and ``cross" modes (or, equivalently, the left-hand and right-hand polarization modes), are always generated in the theories including the scalar or vector fields, although in general modified theories (including the Chern-Simons gravities \cite{Li:2022grj}, the scalar-tensor gravities \cite{brans-dicke,smg}, and Einstein-Aether gravity \cite{einstein-aether} and so on), the amplitudes of these extra modes are relatively smaller than the ``plus" and ``cross" modes. Detecting these extra modes with the laser interferometric detectors provides a model-independent way to distinguish GR and other modified gravities, which is another important method to test the fundamental properties of gravity \cite{wenlinqing,ijmpd2009,niu}.

\section{Applications of the parity-violating gravities to the early universe \label{sec4}}
\renewcommand{\theequation}{4.\arabic{equation}} \setcounter{equation}{0}

In addition to gravitational waves produced by celestial sources, there are also primordial gravitational waves produced by quantum fluctuations in the early Universe. PGWs produce a distinguishable signature in the CMB polarization. In the standard cosmological model, PGWs normally produce autocorrelated TT, EE, and BB power spectra, as well as TE cross-power spectra. The large-scale EB and TB power spectra vanish when the parity of PGWs is conserved. Information from these power spectra can be used to probe the primordial fluctuations. In particular, the TB and EB power spectra are good null tests and can be used to detect the presence of instrumental and/or astrophysical system effects \cite{Yadav:2009za,Chen:2022soq}. Meanwhile, the non-zero TB and EB spectra of PGWs imply parity-violating of the gravitational sector, and their precise measurement is also of great significance in testing for parity-violating interactions \cite{Peng:2022ttg,Qiao:2019hkz,Wang:2012fi, Zhu:2013fja,Zhu:2022dfq}.

\subsection{ Polarized primordial gravitational waves in the ghost-free parity-violating gravity }
\renewcommand{\theequation}{4.1.\arabic{equation}} \setcounter{equation}{0}

For primordial gravitational wave studies, we consider metric perturbations as presented in the first subsection of the previous section for a flat FWR universe. From the action (\ref{xaction}) of this parity-violating gravity, it can be found that the parity-violating terms have no effect on the background evolution. In general, it is assumed that the universe is dominated by scalar field $\phi$ which plays the inflaton field role during the slow-roll inflation. In this case, the Friedmann equation, which governs the background evolution, takes exactly the same form as that in GR, i.e.,
\bqn
H^2 = \frac{8 \pi G}{3} \rho,
\eqn
where $H$ denotes the Hubble parameter during the inflationary stage, and the energy density of the scalar field is $\rho=\frac{1}{2}\dot \phi^2 +V(\phi)$. The evolution of the scalar field $\phi$ is also the same as that in GR,
\bqn
\ddot \phi + 3 H \dot \phi + \frac{dV(\phi)}{d\phi}=0.
\eqn
Typically, in standard slow-roll inflation, the scalar field is assumed to satisfy the slow-roll condition
\bqn\lb{Con}
|\ddot \phi| \ll |3 H \dot \phi|, \;\;\; |\dot\phi^2| \ll V(\phi).
\eqn
With the above slow-roll conditions, it is convenient to define the following Hubble slow-roll parameters,
\bqn
\epsilon_1 = - \frac{\dot H}{H^2},\;\; \epsilon_2 = \frac{d \ln \epsilon_1}{d \ln a}, \;\; \epsilon_3 =  \frac{d \ln \epsilon_2}{d \ln a}.
\eqn

\subsubsection{ The analytical solution of the motion equation for PGWs }
Primordial gravitational waves are the tensor perturbations of the homogeneous and isotropic background, and the equation of motion has been given in Eq.(\ref{eom}). In order to calculate the power spectrum of the primordial gravitational waves, the processing of the equation of motion (\ref{eom}) is different from the previous section, that is, the analytical solution to the equations of motion is given by using the uniform asymptotic approximation method \cite{Qiao:2019hkz}.

In order to facilitate the processing of the equation of motion, a new variable needs to be defined as $u_A\equiv zh_A$, and rewrite Eq.(\ref{eom}) as,
\bqn\lb{Eq}
u''_A+\left[(1+\mu_A)k^2-{z''}/{z}\right]u_A=0,
\eqn
where $z=a\left(1-{c_1k\rho_A}/{(aM_{\rm PV})}\right)^{1/2}$.
Several basic assumptions are worth noting here: a) the PGWs propagate during the inflationary stage; {b) the background evolution during the inflation must satisfy the slow-roll condition (\ref{Con})}; c) the deviations from GR arising from the parity violation are small.
 Considering these factors, the effective time-dependent mass term $z''/z$ in Eq.(\ref{Eq}) is expanded in terms of the slow-roll parameter as
\bqn\lb{z}
\frac{z''}{z}&=&\frac{a''}{a}+\frac{1}{2}\frac{\left(\frac{a''}{a}c_1-c_1''\right)k\rho_A/aM_{\rm PV}}{1-c_1k\rho_A/aM_{\rm PV}} +\frac{1}{4}\left[\frac{(c_1\mathcal{H}-c_1')k\rho_A/aM_{\rm PV}}{1-c_1k\rho_A/aM_{\rm PV}}\right]^2\nb\\
&\simeq&\frac{v_t^2-\frac{1}{4}}{\tau^2}-\rho_A\frac{k}{\tau}c_1\epsilon_*,
\eqn
where
\bqn
v_t \simeq \frac{3}{2}+3 \epsilon_1+4\epsilon_1^2+4\epsilon_1\epsilon_2+\mathcal{O}(\epsilon^3),
\eqn
and $\epsilon_* \equiv {H}/{M_{\rm PV}}$ denotes the ratio between the energy scale of inflation and the characteristic energy scale of the parity violation, which determines the magnitude of the corrections to GR.

Similarly, the parameter $\mu_A$ that modifies the dispersion relation of the tensor mode can also be expressed in the form of slow-roll parameter as
\bqn\lb{u}
\mu_A &=& \frac{\rho_A k (c_1 -c_2)/(a M_{\rm PV})}{1- \rho_A k c_1/(a M_{\rm PV})}\nb\\
&\simeq&-\rho_Ak\tau(c_1-c_2)\epsilon_*.
\eqn
It is worth noting that in order to obtain the above expansion, the following relation is used,
\bqn
a&=&-\frac{1}{\tau H}\left(1+\epsilon_1+\epsilon^2_1+\epsilon_1\epsilon_2\right)+\mathcal{O}(\epsilon^3).
\eqn

Combining the expressions for $z''/z$ and $\mu_A$, the equation of motion in Eq.(\ref{Eq}) can be transformed into the form
\bqn\lb{zz}
&&u''_A+\Bigg\{\big[1-\rho_Ak\tau(c_1-c_2)\epsilon_*\big]k^2-\frac{v_t^2-\frac{1}{4}}{\tau^2} +\rho_A\frac{k}{\tau}c_1\epsilon_*\Bigg\}u_A=0.
\eqn
After some tricksy calculation, a more familiar ordinary differential equation is obtained. Since this equation reduces to the same form as in CS gravity when $c_1=c_2$, it has an exact solution in terms of the confluent hypergeometric functions given in Ref. \cite{Bartolo:2017szm}. However, in this case with coefficients $c_1\neq c_2$, there is no known exact solution to the equation, which urges a solution. In general, the most widely considered method is the WKB approximation, in which the satisfaction of WKB condition is claimed throughout the process. Nevertheless, there are situations in which the WKB conditions may be broken or not entirely satisfied \cite{Zhu:2013upa}.


Here applying the uniform asymptotic approximation \cite{Zhu:2013upa}, the final expression for the approximate solution through tedious calculations is given in terms of Airy type functions as \cite{Qiao:2019hkz}
\bqn\lb{ALY}
u_A=\alpha_0\left(\frac{\xi}{g(y)}\right)^{1/4}{\rm Ai}(\xi)+\beta_0\left(\frac{\xi}{g(y)}\right)^{1/4}{\rm Bi}(\xi),
\eqn
where $\rm{Ai}(\xi)$ and $\rm{Bi}(\xi)$ are the Airy functions, $\alpha_0$ and $\beta_0$ are two integration constants, $\xi$ is the function of $y$ and their expressions are given as follows:
\bqn
\alpha_0&=&\sqrt{\frac{\pi}{2k}}e^{i\frac{\pi}{4}},~~~
\beta_0=i\sqrt{\frac{\pi}{2k}}e^{i\frac{\pi}{4}},\\
\xi(y) &=&
\begin{cases}
\left(-\frac{3}{2}\int^y_{y_0}\sqrt{g(y')}dy'\right)^{2/3} ,\;  & y\leq y_0,\\
-\left(\frac{3}{2}\int^y_{y_0}\sqrt{-g(y')}dy'\right)^{2/3} ,\; & y\geq y_0.\\
\end{cases}
\eqn
At this point, only the specific forms of the Airy functions in the expression for the approximate solution have not been determined.
It is clear from the expressions for  $\xi$ above that the value of  $\xi$ depends on $y$, while the value of  $\xi$ influences the choice of forms of the Airy functions. Consequently, the choice of explicit forms for the Airy functions is left to be determined later for the specific condition required for the calculation.

\subsubsection{ Power spectra of PGWs }

With the approximate solution of PGWs derived above, the associated power spectrum $\mathcal{P}^{L,R}_{\rm T}$ can be computed in the limit $y \to 0 $ by
\bqn
\mathcal{P}_{\rm T}^{\rm L} = \frac{2 k^3}{\pi^2} \left|\frac{u_k^{\rm L}(y)}{z}\right|^2,
\mathcal{P}_{\rm T}^{\rm R} = \frac{2 k^3}{\pi^2} \left|\frac{u_k^{\rm R}(y)}{z}\right|^2.
\eqn
Here, before calculating the specific power spectrum, the form of the required Airy functions needs to be determined. When $y\to 0^+$, the parameter $\xi(y)$ becomes very large and positive, and the Airy function allows the following asymptotic forms
\bqn\lb{AB}
{\rm Ai}(x)&=&\frac{1}{2\pi^{1/2}x^{1/4}}\exp\left(-\frac{2}{3}x^{3/2}\right),\\
{\rm Bi}(x)&=&-\frac{1}{\pi^{1/2}x^{1/4}}\exp\left(\frac{2}{3}x^{3/2}\right).
\eqn
From the above two expressions for the Airy function, it is clear that only the growing mode of $u_k^A(y)$ is relevant under the limit $y \to 0 $, so we have
\bqn
u_k^A(y)&\approx&\beta_0\left(\frac{1}{\pi^2g(y)}\right)^{1/4}\exp\left(\int^{y_0}_ydy\sqrt{g(y)}\right)\nb\\
&=&i\frac{1}{\sqrt{2k}}\left(\frac{1}{g(y)}\right)^{1/4}\exp\left(\int^{y_0}_ydy\sqrt{g(y)}\right).\nb\\
\eqn
The power spectra of PGWs are then given by
\bqn\lb{PS}
\mathcal{P}^A_{\rm T}
&=&\frac{k^2}{\pi^2}\frac{1}{z^2}\frac{y}{v_t}\exp\left(2\int^{y_0}_ydy\sqrt{g(y)}\right)\nb\\
&\simeq&18\frac{H^2}{\pi^2 e^{3}}e^{\frac{\pi\rho_A\epsilon_*}{16}(9c_2-c_1)}\nb\\
&\simeq&18\frac{H^2}{\pi^2 e^{3}}\left[1+\frac{\pi\rho_A}{16}\mathcal{M}\epsilon_*+\frac{\pi^2 }{2\times16^2}\mathcal{M}^2\epsilon_*^2+\mathcal{O}(\epsilon_*)^3\right],\nb\\
\eqn
where $\mathcal{M}  \equiv 9c_2-c_1$ and
\bqn
\frac{9c_2-c_1}{M_{\rm PV}}&=& 8\dot{\vartheta}+4 \dot{a_1}\dot{\phi}^2 +8 a_1\dot{\phi}\ddot{\phi} - 8a_1 H \dot{\phi}^2- 16\dot{a_2}\dot{\phi}^2
- 32a_2\dot{\phi}\ddot{\phi} + 10\dot{a_3}\dot{\phi}^2 + 20a_3\dot{\phi}\ddot{\phi} -  4a_3 H \dot{\phi}^2
-64\dot{a_4}\dot{\phi}^2-128a_4\dot{\phi}\ddot{\phi}  \nb\\
&& +2 b_1\dot{\phi}^3-4b_2\left(H\dot{\phi}^2-\dot{\phi}\ddot{\phi}\right)
-2b_3 \dot{\phi}^3\ddot{\phi}  - 2b_4 \dot{\phi}^3\ddot{\phi}
+2b_5 H \dot{\phi}^4-2b_7\dot{\phi}^3\ddot{\phi} -6b_7\dot{\phi}^4H .
\eqn
It is clear that the power spectra can be modified due to the presence of a parity-violating term in the action (\ref{xaction}). As expected, it can be checked that the standard GR results are recovered when $\mathcal{M}\epsilon_*= 0$. Therefore, the power spectra in (\ref{PS}) is rewritten as
\bqn
\mathcal{P}^A_{\rm T}= \frac{\mathcal{P}^{\rm GR}_{\rm T}}{2}\left[1+\frac{\pi\rho_A}{16}\mathcal{M}\epsilon_*+\frac{\pi^2\rho^2_A}{2\times16^2}\mathcal{M}^2\epsilon_*^2+\mathcal{O}(\epsilon_*)^2\right],\nb\\
\eqn
where
 \bqn
\mathcal{P}^{\rm GR}_{\rm T}=\frac{2 k^3}{\pi^2}\left(\left|\frac{u_k^{\rm L}(y)}{z}\right|^2+\left|\frac{u_k^{\rm R}(y)}{z}\right|^2\right)
\eqn
denotes the standard nearly scale invariant power-law spectrum calculated by uniform asymptotic approximation in the framework of GR \cite{Zhu:2013upa}. For the two circular polarization modes, namely $A={\rm R}$ and $A={\rm L}$, the spectra $\mathcal{P}^{\rm GR}_{\rm T}$ have the exactly same form. The quantity $\mathcal{M}$ depends on the coefficients $\vartheta$, $a_{\mathcal{A}}$ and $b_{\mathcal{A}}$, as well as the evolution of the scalar field.
It is worth noting that for positive values of $\mathcal{M}$, parity violation has a tendency to enhance (suppress) the power spectra of the right (left) handed mode. During the slow-roll inflation, the scalar field is slow-rolling, which satisfies the slow-roll conditions (\ref{Con}).  With this condition, the quantities $c_1$ and $c_2$ are assumed to be slowly varying during the expansion of the universe, which can be approximately treated as constants during the slow-roll inflation. In the expression of $9c_2-c_1$, we observe that it contains the terms with $\vartheta, a_{\mathcal{A}}, b_{\mathcal{A}}$ and their derivatives with respect to $\phi$. Considering the scalar field $\phi$ with the slow-roll condition (\ref{Con}), the leading contribution to $9c_2-c_1$ reads
\bqn
\frac{9c_2-c_1}{M_{\rm PV}}\simeq 8 \dot \vartheta -8 (a_1+ \frac{a_3}{2}+ \frac{b_2}{2} ) H \dot \phi^2.
\eqn
 Therefore, the leading contribution to the power spectrum of PGWs depends only on the coefficients $\dot \vartheta, a_1, a_3$ and $b_2$.

\subsubsection{ The circular polarization and detectability }

A quantity that can be directly detected is the degree of circular polarization of the PGWs, which is defined by the differences in the amplitudes between the two circular polarization states of PGWs as
\bqn
\Pi&\equiv&\frac{\mathcal{P}^{\rm R}_{\rm T}-\mathcal{P}^{\rm L}_{\rm T}}{\mathcal{P}^{\rm R}_{\rm T}+\mathcal{P}^{\rm L}_{\rm T}} \simeq { \frac{\pi}{16} (9c_2-c_1)\epsilon_* } +\mathcal{O}(\epsilon_*^3)\nb\\
&\simeq&\frac{\pi}{2} \dot \vartheta M_{\rm PV}\epsilon_*-\frac{\pi}{2}(a_1+ \frac{a_3}{2}+ \frac{b_2}{2})H \dot \phi^2M_{\rm PV} \epsilon_*+\mathcal{O}(\epsilon_*^3).
\eqn
As expected, when $a_1 = a_3 = b_2 =0$, the above expression of the circular polarization $\Pi$ exactly reduces to that in Chern-Simons gravity \cite{Alexander:2004wk, Satoh:2008ck}. Obviously, under conditions (\ref{Con}), we observe that the degree of the circular polarization $\Pi$ is very small due to the suppressing parameter $\epsilon_*$.

We further analyze this observable $\Pi$, which provides the opportunity to directly detect the chiral asymmetry of gravity by observations \cite{Lue:1998mq, Saito:2007kt, Gluscevic:2010vv}. However, as being pointed out in \cite{Wang:2012fi}, the detectability of the circular polarization of PGWs is sensitive to the values of the tensor-scalar-ratio $r$ and $\Pi$. According to the combination of Planck 2018 data and the BICEP2/Keck Array BK15 data \cite{Planck:2018jri}, $r$ has been tightly constrained as $r \lesssim 0.065$. For this case, in order to detect any signal of parity violation in the forthcoming CMB experiments,  $\Pi$ must be larger than $\mathcal{O}(0.5)$, even in an ideal case with the cosmic variance limit. On the other hand, since the condition $\epsilon_* \ll 1$ is imposed for the original considerations when constructing the theory, the order of magnitude of $\Pi$ is rough $\lesssim \mathcal{O}(0.5)$. For these reasons, we conclude that it seems difficult to detect or efficiently constrain the parity violation effects on the basis solely of two-point statistics from future CMB data.

\subsection{Primordial gravitational waves in parity-violating Ho\v{r}ava-Lifshitz gravity}
\renewcommand{\theequation}{4.2.\arabic{equation}} \setcounter{equation}{0}

We have discussed the corrections to the GWs propagation waveform in parity-violating Ho\v{r}ava-Lifshitz gravity compared to that in GR. Here we continue to introduce the effects of parity violation on PGWs in this gravity.

\subsubsection{ The analytical solution of the motion equation for PGWs }

To facilitate solving Eq. (\ref{HLeq}), define new variables $u_A\equiv \sqrt{\alpha k} a h_A$ and $y\equiv -\alpha k \tau$. Rewrite Eq. (\ref{HLeq}) with these two variables as
\bqn\lb{HLxeq}
u_{A,yy} + (\omega^2_A - 2y^{-2}) u_A = 0,
\eqn
where
\bqn
\omega^2_A = 1+ \rho^A(\delta_1y+ \delta_3 y^3) + \delta_2y^2 + \delta_4y^4,
\eqn
with $\delta_1 \equiv -2(\alpha_1/\alpha^3) \epsilon_*,~\delta_2 \equiv (\hat{\gamma}_3/\alpha^4) \epsilon_*^2,~\delta_3 \equiv (\alpha_2/\alpha^5) \epsilon_*^3,~ \delta_4 \equiv (\hat{\gamma}/\alpha^6) \epsilon_*^4$ and $\epsilon_* \equiv H/M_{\rm PV} \ll 1$. Note that the unitarity of the theory in the UV requires $\hat{\gamma}_5 > 0$, while a healthy IR limit requires $\alpha^2 \rightarrow 1$. Thus, without loss of generality, $\alpha = 1$, or equivalently $\alpha_0= 0$, is set in the following. However, $\alpha_1$, $\alpha_2$ and $\hat{\gamma}_3$ have no such restrictions, provided that $\omega^2_A > 0$ holds.

In order to study the power spectrum of PGWs, it is first necessary to solve Eq. (\ref{HLxeq}). The solution of Eq. (\ref{HLxeq}) is related to the choice of parameters in the dispersion relation $\omega^2_A$. Through meticulous analysis and tedious calculations in Ref. \cite{Wang:2012fi}, the mode functions $u_R = \sqrt{\alpha k} v_R $ and $u_L = \sqrt{\alpha k} v_L$ of Eq. (\ref{HLxeq}) are given by
\bqn
v_R &=&\begin{cases}
\f{1}{\sqrt{2\omega_R} }e^{-i\int ^{\tau}_{\tau_i} \omega_R(k,\tau')d\tau' },\;  & \omega_R > H,\\
D_+ a (\tau) + D_{-} a(\tau) \int^{\tau}_{\tau_3}\f{d\tau'}{a^2(\tau')} ,\; & \omega_R<H,\\
\end{cases}\\
v_L &=&
\begin{cases}
\f{1}{\sqrt{2\omega_L} }e^{-i\int ^{\tau}_{\tau_i} \omega_L(k,\tau')d\tau' },\;  & \tau \in (\tau_i, \tau_1),\\
C_+ a (\tau) + C_{-} a(\tau) \int^{\tau}_{\tau_1}\f{d\tau'}{a^2(\tau')} ,\; &  \tau \in (\tau_1, \tau_2),\\
\f{\alpha_k e ^{-i\Theta^L_2(k,\tau)} + \beta_k e^{i\Theta^L_2(k,\tau)}}{\sqrt{2\omega_L(k,\tau)}} ,\; &  \tau \in (\tau_2, \tau_3),\\
D_+ a (\tau) + D_{-} a(\tau) \int^{\tau}_{\tau_3}\f{d\tau'}{a^2(\tau')} ,\; &  \tau \in (\tau_3, 0),
\end{cases}
\eqn
where $\Theta^A_n(k,\tau) \equiv \int^{\tau}_{\tau_n} \omega_A(k,\tau')d\tau'$. The coefficients $C_{\pm}(k), ~D_{\pm}(k),~\alpha_k,~\beta_k$ are uniquely determined by requiring that $v^{R,L}_k$ and its first-order time derivative be continuous across the boundaries that separate these regions.

\subsubsection{ The power spectrum and circular polarization }

In order to investigate the effect of the parity-violating terms, two representative cases are considered in Ref. \cite{Wang:2012fi}: (i) $\delta_2 = \delta_3 = 0$; (ii) $\delta_1 = \delta_2 = 0$. In the former, the power spectrum of PGWs and the circular polarization are given by,
 \bqn
\mathcal{P}_{\rm T} &=& \frac{ k^3\left( |h_R|^2 + |h_L|^2 \right)}{(2\pi)^2} = \f{H^2}{4\pi^2} \left( 1+ 21\alpha^2_1 \epsilon_*^2 + \mathcal{O}(\epsilon^3_*) \right),\\
\Pi &= & \f{ |h_R|^2 - |h_L|^2}{ |h_R|^2 + |h_L|^2} = 3\alpha_1\epsilon_* + \left( 17\alpha^3_1 - 3\alpha_2  \right) \epsilon_*^3/2  + \mathcal{O}(\epsilon^5_*) .
\eqn
In the latter case, the power spectrum and circular polarisation of the PGWs are given by
\bqn
\mathcal{P}_{\rm T} &=& \f{H^2}{4\pi^2} \left[ 1+ \Delta^L_k - 3\alpha_1 \Delta^L_k\epsilon_* + \f{21}{2}(1+ \Delta^L_k) \alpha^2_1 \epsilon^2_*+ \mathcal{O}(\epsilon^3_*) \right],\\
\Pi &= &  - \f{\Delta^L_k}{1+ \Delta^L_k}+ \f{3(1+ 2\Delta^L_k)\alpha_1}{(1+ \Delta^L_k)^2}\epsilon_* + \f{9\Delta^L_k(1+ 2\Delta^L_k)\alpha^2_1}{(1+ \Delta^L_k)^3}\epsilon^2_* + \mathcal{O}(\epsilon^3_*) ,
\eqn
where $\Delta^A_k \equiv |\beta^A_k|^2 + {\rm Re}\left( \alpha^A_k\beta^{A_*}_k e^{-2i\Theta^A_{23}} \right)$, and $\Theta^A = \Theta^A_n(k,\tau_m)$.

In addition to the above two specific cases, when $\delta_2 \neq 0$, there is another possibility in which $\omega_{R, L}=H $ has three real positive roots. The power spectrum and circular polarisation of the PGWs are given by
\bqn
\mathcal{P}_{\rm T} &=& \f{H^2}{4\pi^2} \left[ 1+ \Delta^L_+ - 3\alpha_1 \Delta^-_k\epsilon_* + \f{3}{2}(7\alpha^2_1 - \hat{\gamma}_3)(1+ \Delta^+_k) \epsilon^2_*+ \mathcal{O}(\epsilon^3_*) \right],\\
\Pi &= &  - \f{\Delta^-_k}{1+ \Delta^+_k}+ \f{3\alpha_1(1+ 2\Delta^R_k)(1+ 2\Delta^L_k)}{(1+ \Delta^+_k)^2}\epsilon_* + \f{9\alpha^2_1 \Delta^-_k(1+ 2\Delta^R_k)(1+ 2\Delta^L_k)}{(1+ \Delta^+_k)^3}\epsilon^2_* + \mathcal{O}(\epsilon^3_*) ,
\eqn
where $\Delta^{\pm}_k \equiv \Delta^L_k \pm \Delta^R_k $.
The power spectra of PGWs with parity violation corrections in parity-violating HL gravity for the cases with different parameters are given above. Each of these power spectra corresponds to a non-zero circular polarization $\Pi$.
These three circular polarisations $\Pi$ have been analyzed in Ref. \cite{Wang:2012fi}, and a large $\Pi$ is possible in all three cases, i.e. reaching detectable probability. In the analysis of polarized PGWs for the ghost-free parity-violating gravity, it has been shown that the detectable quantity $\Pi$ must require $\Pi > \mathcal{O}(0.5)$  in order to be experimentally detectable. Thus, the parity-violating HL gravity produces a stronger correction in the power spectrum of PGWs for parity violation compared to the ghost-free parity-violating gravity. The degree of circular polarization of PGWs can be detected in future three-point correlation function statistics of CMB data.

\subsection{Intercomparison of parity-violating gravities}

This section describes the application of these parity-violating gravities in the early universe.  The effect of the parity-violating terms on the circular polarization of the PGWs is presented in detail for the ghost-free parity-violating gravity and the parity-violating HL gravity. It is shown that with the presence of the parity violation, the power spectra of PGWs are modified and the degree of circular polarization becomes nonzero. The circular polarization generated in the ghost-free parity-violating theory of gravity is quite small, which is suppressed by the energy scale of parity violation of the theory, and it is difficult to detect by using the power spectra of future CMB data. However, the circular polarization produced in parity-violating HL gravity is large and detectable using future CMB data's power spectrum. It is necessary to mention that, the primordial gravitational waves is detectable, only if the energy scale of inflation is around the Grand Unified Theory (GUT) scale of $10^{16}$GeV. If the parity-violating effects of primordial gravitational waves will be detected in future, the value of $\Pi$ must be $\mathcal{O}(1)$, which indicates that the energy scale of parity violation in gravity is also round the similar scale. Therefore, the detection of primordial gravitational waves with cosmic microwave background radiation provides an excellent opportunity to test the parity symmetry of gravity at the GUT scale.

{In Nieh-Yan modified teleparallel gravity and parity-violating symmetric teleparallel gravity, in order to explore the mechanism of the primary perturbation generation, the quadratic perturbation of the action is given respectively \cite{Li:2020xjt,Li:2021mdp}. In the Nieh-Yan modified teleparallel gravity, the quadratic action on the scalar perturbation clearly shows that there is only one dynamic scalar degree of freedom, although two scalar fields were introduced at the beginning. This may be useful for searching for new mechanisms of generating primordial perturbations. The quadratic action on the tensor perturbation shows that only velocity birefringence and not the amplitude birefringence of GW is produced. The parity-violating term in parity-violating symmetric teleparallel gravity does not influence the scalar perturbations. Vector perturbations exhibit both velocity and amplitude birefringence effects in this gravity, where one of the vector modes is a ghost at high momentum scales, which would lead to vacuum instabilities in the quantum theory of cosmic perturbations. The quadratic action on the tensor perturbation also shows only velocity birefringence, which is the same as the correction for Nieh-Yan modified teleparallel gravity.}

\section{Conclusions and Discussions\label{sec5}}
\renewcommand{\theequation}{5.\arabic{equation}} \setcounter{equation}{0}

In this article, we review the applications of different parity-violating gravities. We discuss the application of parity-violating gravities to isolated sources. The presence of parity violation in parity-violating HL gravity, Nieh-Yan modified teleparallel gravity and parity-violating symmetric teleparallel gravity induces only a velocity birefringence, which is the opposite of the amplitude birefringence effect present in CS gravity. However, in ghost-free parity-violating gravity, the presence of the parity violation induces both velocity and amplitude birefringence. These effects are fully consistent with the existence of both ways of parity violation, where in parity-violating HL gravity, Nieh-Yan modified teleparallel gravity and parity-violating symmetric teleparallel gravity the parity-violating terms indeed correct only the dispersion relation in the GWs propagation equation; in ghost-free parity-violating gravity the parity-violating terms change both the dispersion relation and the friction term in the GWs propagation equation. Based on these corrected waveforms, the parity-violating gravities are examined separately and analyzed using the currently observed GWs events. The results present that there is no significant parity violation in parity-violating gravities. Meanwhile, the parity-violating gravities are given separately with the latest constraints on the energy scale: $M_{\rm PV} > 0.09 \rm GeV$ for the ghost-free parity-violating gravity and $M_{\rm PV}<6.5 \times 10^{-42} \rm GeV$ for the Nieh-Yan modified teleparallel gravity and the parity-violating symmetric teleparallel gravity.

In addition, we present the application of parity-violating gravities to the early universe. In the ghost-free parity-violating gravity, it is shown that with the presence of the parity violation, the power spectra of PGWs are slightly modified and the degree of circular polarization becomes nonzero. The circular polarization generated in the ghost-free parity-violating theory of gravity is quite small, which is suppressed by the energy scale of parity violation of the theory. In order to detect any parity-violating signals in the upcoming CMB experiments, $\Pi$ must be larger than $\mathcal{O}(0.5)$. The analysis shows that $\Pi$ is rough $\lesssim \mathcal{O}(0.5)$ in ghost-free parity-violating gravity, which is difficult to detect by using the power spectra of future CMB data. Similarly, the power spectrum of PGWs in parity-violating HL gravity is also corrected due to the presence of the parity violation. However, the resulting non-zero circular polarization $\Pi$ might be large enough that it can be detected in future three-point correlation function statistics of CMB data.
{

\section*{Acknowledgements}

We appreciate the helpful discussion with Mingzhe Li and Dehao Zhao. This work is supported by the National Key R\&D Program of China Grant No. 2022YFC2204602 and 2021YFC2203102, NSFC No. 12273035 and 11903030 the Fundamental Research Funds for the Central Universities under Grant No. WK2030000036 and WK3440000004, and the science research grants from the China Manned Space Project with No.CMS-CSST-2021-B01. T.Z. is supported in part by the National Key Research and Development Program of China under Grant No.2020YFC2201503, the National Natural Science Foundation of China under Grant No. 12275238 and No. 11675143, the Zhejiang Provincial Natural Science Foundation of China under Grant No. LR21A050001 and LY20A050002,  and the Fundamental Research Funds for the Provincial Universities of Zhejiang in China under Grant No. RF-A2019015.

\end{document}